\documentclass[runningheads]{llncs}
\usepackage{graphicx}
\usepackage{multirow}
\usepackage{booktabs}
\usepackage{soul}
\usepackage{algorithm}
\usepackage[noend]{algpseudocode}
\usepackage{wrapfig}
\usepackage{amsfonts}
\usepackage{arydshln}
\usepackage{colortbl}
\usepackage{hhline}

\algnewcommand\algorithmicforeach{\textbf{for each}}
\algdef{S}[FOR]{ForEach}[1]{\algorithmicforeach\ #1\ \algorithmicdo}
\newcommand{\algorithmicbreak}{\textbf{break}}
\newcommand{\BREAK}{\State \algorithmicbreak}

\begin{document}
\title{I Will Survive: An Online Conformance Checking Algorithm Using Decay Time}

\author{Kristo Raun and Ahmed Awad}
\institute{University of Tartu, Tartu, Estonia \email{\{kristo.raun,ahmed.awad\}@ut.ee}}
\titlerunning{I Will Survive: An Online Conformance Checking Algorithm}
\maketitle              
\begin{abstract}
Process executions in organizations generate a large variety of data. Process mining is a data-driven analytical approach for analyzing this data from a business process point of view. Notably, the activities executed in organizations do not always conform to how the processes have been modeled. Online conformance checking deals with finding discrepancies between real-life and modeled behavior on data streams.   
The current state-of-the-art output of online conformance checking is a prefix-alignment, which is used for pinpointing the exact deviations in terms of the trace and the model while accommodating a trace's unknown termination in a streaming setting. However, producing prefix-alignments entails a state space search to find the shortest path from a common start state to a common end state between the trace and the model. This is computationally expensive and makes the method infeasible in an online setting.

Previously, the trie data structure has been shown to be efficient for constructing alignments, utilizing a proxy log that represents the process model in a finite way. This paper introduces a new approximate algorithm (IWS) on top of the trie for online conformance checking. The algorithm is shown to be fast, memory-efficient, and able to output both a prefix and a complete alignment event-by-event while keeping track of previously seen cases and their state. Comparative analysis against the current state-of-the-art algorithm for finding prefix-alignments shows that the IWS algorithm achieves, in some cases, an order of magnitude faster execution time while having a smaller error cost. In extreme cases, the IWS finds prefix-alignments roughly three orders of magnitude faster than the current state-of-the-art. The IWS algorithm includes a discounted decay time setting for more efficient memory usage and a look-ahead limit for improving computation time. Finally, the algorithm is stress tested for performance using a simulation of high-traffic event streams.

\keywords{Process mining, Conformance checking, Event streams, Approximate methods}

\end{abstract}
\section{Introduction} \label{sec:introduction}


Process mining is a data-driven approach for analyzing process execution data. This data is commonly collected in so-called process execution event logs. In its simplest form, an execution log is a sequence of events. Each event is commonly characterized by at least three attributes. The case identifier, a reference to the process instance in which it was generated, the activity identifier, a reference to the activity that has been executed, and a timestamp indicating the event time. An example event log is shown in Table~\ref{tab:event_log_example}. While further data attributes may exist, they are not within the scope of this paper. Events having the same case identifiers are called a trace and they represent the sequence of actions that occurred within the context of this process instance.

\begin{table}[]
\begin{tabular}{@{}ccc@{}}
\toprule
\textbf{Case id} & \textbf{Activity} & \textbf{Timestamp} \\ \midrule
1                & a                 & 2022-08-01 15:00   \\
1                & b                 & 2022-08-01 15:02   \\
2                & a                 & 2022-08-01 15:03   \\
2                & b                 & 2022-08-01 15:06   \\
1                & c                 & 2022-08-01 15:06   \\ \bottomrule
\end{tabular}
\caption{A simple event log showing the case identifier, executed activity, and execution timestamp.}
\label{tab:event_log_example}
\end{table}

Process mining is commonly divided into three main areas: process discovery~\cite{firstPRocessMiningPaper98}, conformance checking~\cite{carmona2018conformance}, and process enhancement~\cite{processMiningBook}. Process discovery deals with \emph{discovering} the underlying process model from execution logs. Conformance checking looks at how well the process executions and the process model align. The goal of process enhancement is to improve or extend the current process model.

\begin{figure}[h]
	\centering
	\includegraphics[width=0.6\linewidth]{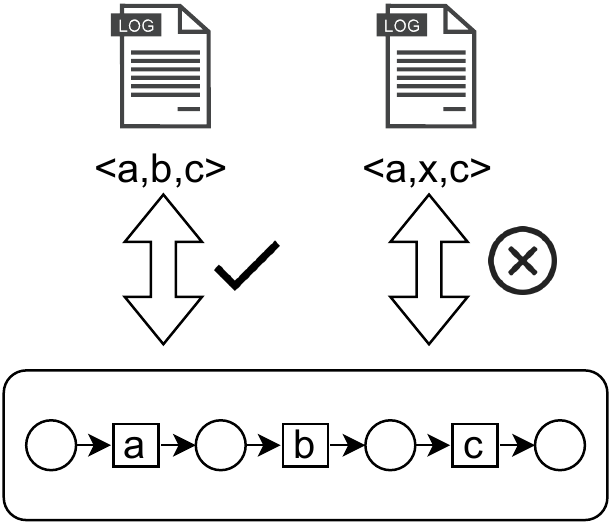}
	\caption{Conformance checking is the comparison of process executions, i.e. an event log or event stream, and a process model (here presented as a Petri net).}
	\label{fig:confchecking}
\end{figure}

Conformance checking assumes that we have knowledge of how the world should work - i.e., we have a process model - and we have examples of how the world is actually working - i.e., process traces. We then compare the traces to the model in order to analyze the conformance of the process (Figure~\ref{fig:confchecking}). Non-conformance may not necessarily indicate wrong activities in the execution of the process - it may also be a sign of possible process enhancement. Regardless, it is important to be able to find the discrepancies that exist in modeled and real behavior. 

The research on conformance checking started mostly in the offline setting. The workflow is that event data for a specific time range is extracted from the business systems and a conformance artifact is produced on top of this data. Various conformance checking paradigms have been suggested, but the commonly accepted state-of-the-art for the past decade has been an \emph{alignment} between the trace and the model\cite{adriansyah2014aligning}. 

An organization can have thousands of ongoing process executions at the same time. Thus, an output from the business systems should either remove unfinished traces or take into account that the analysis will include both finished and unfinished traces. Otherwise, the analysis penalizes the unfinished cases as they are considered non-conforming to the process model. Furthermore, an analysis of past data may quickly lose its value, as usually deviations need to be discovered and acted upon in a timely manner. Such observations have paved the way for online conformance checking. 

In online conformance checking, the conformance checking is done on event streams, rather than static event logs. The underlying goal is the same - finding discrepancies between modeled and real-life behavior. However, in an online setting, the termination of a single process execution is unknown and the event stream is unbounded. Thus, fast-paced event streams require a computationally efficient algorithm to keep up with the incoming data. While efficient algorithms exist for online conformance checking, they do not use alignments as their output\cite{burattin2018online}. At the same time, methods using prefix-alignments have been introduced for online conformance checking\cite{van2019online,schuster2020online}, but their computational complexity hinders their applicability in real-life settings.

This paper attempts to bridge this gap by introducing a new efficient algorithm for online conformance checking. The algorithm outputs prefix-alignments with comparable error costs to the state-of-the-art while improving the computation time by a noticeable extent. The paper is structured as follows: in Section~\ref{sec:background} a theoretical background is given. In Section~\ref{sec:related} related works are presented. Section~\ref{sec:approach} introduces the approach and the algorithm. Section~\ref{sec:results} compares the algorithm to the state-of-the-art in terms of cost deviations and execution time. Also, a stress test under a fast-paced stream is performed to validate the algorithm's applicability in real-life settings. Finally, Section~\ref{sec:conclusion} summarizes the work and provides venues for future research.

\section{Background} \label{sec:background}

In this section, we introduce the main components necessary for understanding the theoretical background of the introduced approach. We describe process models, event logs, the trie data structure, conformance checking, and the trie data structure adaption to the context of (online) conformance checking. Furthermore, conformance checking is discussed in terms of alignments and why in online conformance checking it is sensible to use \emph{prefix-}alignments.

\subsection{Process Models}
\label{sec:process:models}
A process model defines which sequences of activity executions are considered to be valid. There are many notations to model business processes varying in their richness and formal semantics. For this paper, we adopt a special case of a Petri net called a Workflow net~\cite{WFNet97} where there is a single source place $i$, and a single sink place $o$, and any other node, i.e., either a place or a transition, is on a path from the source place to the sink place. In other words, if we add another transition $t$ to the net with one arc from $o$ to $t$ and another arc from $t$ to $i$, the resulting Petri net forms a single strongly connected component. Workflow nets follow the standard semantics of transitions enablement and firing as ordinary Place/Transition Petri nets~\cite{petriNets2019}. 

\begin{figure}[h]
	\vspace{-.5em}
	\centering
	\includegraphics[width=0.75\linewidth]{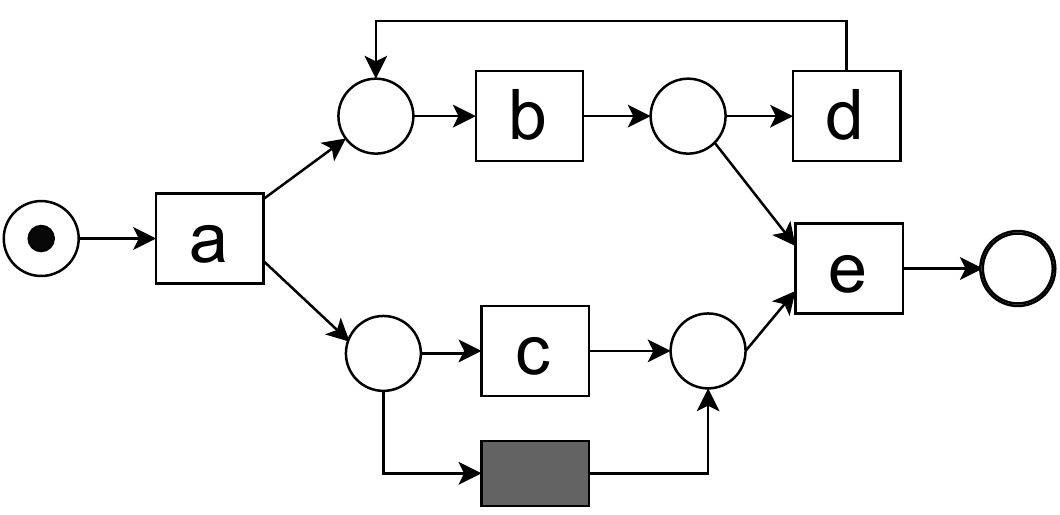}
	\vspace{-.5em}
	\caption{Example process model for conformance checking}
	\label{fig:petri:net}
	\vspace{-.5em}
\end{figure}

The workflow net in Figure~\ref{fig:petri:net} serves as our running example having five labeled transitions: $a,b,c,d,e$. There is also a silent $\tau$ transition, colored in grey. Silent transitions are transitions that cannot be observed during the execution of a process model. For example, in Figure~\ref{fig:petri:net}, the $\tau$ transition allows for a skip on $c$, but there is no labeled activity associated with skipping $c$ that could be shown on the model.

When workflow nets are enacted, one can observe sequences of labeled transitions based on firing sequences. As such, its behavior $M$ may also be represented by a set of sequences of activities. $M$ is infinite when the model has loops because a loop can unfold an unlimited number of times. 
The sequence of fired transitions is called an execution sequence. An execution sequence $\pi \in M$ starts from a transition enabled by the source place marking and ends with a transition that marks the sink place after its firing. A prefix of an execution sequence ${}_i\pi$ indicates the execution until $i$-th position in $M$.
An instance of an execution sequence is shown as $\pi=\langle a, b,c, e\rangle$, representing the execution path followed by executing, a, b, c, and e transitions from the workflow net in Figure~\ref{fig:petri:net}. Another execution sequence $\pi=\langle a, b, d, c, b, d, b, d, e\rangle$ has the loop around transitions $b$ and $d$ executed two times. 


\subsection{Event Logs/Streams}
\label{sec:event:logs:Streams}
Data about the execution of a process is commonly represented as an \emph{event 
log}~\cite{processMiningBook}. It contains \emph{traces} that denote single executions of the process, 
each trace being a sequence of events. An \emph{event}, in turn, represents the execution of an activity of the process. Traces representing distinct process 
executions built of events that induce the same sequence of activity 
executions are said to be of the same \emph{trace variant}. Commonly, an \emph{event} contains the \emph{case ID} for assigning an event to a particular process execution, an \emph{activity} indicating the label of the executed event, and the execution \emph{timestamp} of the event. 

While events and traces may be assigned further information about the context 
of the process execution, such as data payload or resource information, a relatively simple 
model for event logs is sufficient for the context of this paper. Specifically, 
with $\mathcal{A}$ as a universe of activities, we model a trace $\sigma$ as a 
finite sequence of activities, $\sigma =\langle a_1,\ldots, a_n \rangle \in 
\mathcal{A}^*$. We use the notation $\sigma(i)$ for the activity at the $i$-th 
position of $\sigma$. $\sigma^\prime \subseteq \sigma$ indicates a subsequence of events belonging to a trace. 

We can observe the similarity between observed traces in actual process execution and firing sequences that can be obtained from workflow nets. A major difference is contextual execution information like the timestamp of executing an activity. In the context of this work, actual timestamp values are not relevant and they are used mainly to reason about the arrival order of events.

In an online setting, conformance checking tools usually observe a subsequence of the trace. For example, a \emph{prefix} of the trace indicates that the process execution may not have concluded at the time of observation. An infix of the trace represents the case when streaming execution started before the conformance checker is up and ready to receive events - also called \emph{warm starting}. In the rest of this paper, the analysis will be limited to the prefix, as the prefix-alignments have been thus far the artifact produced by the state-of-the-art methods.

\subsection{Trie}
\begin{figure}[h!]
    \centering
    \includegraphics[width=0.7\linewidth]{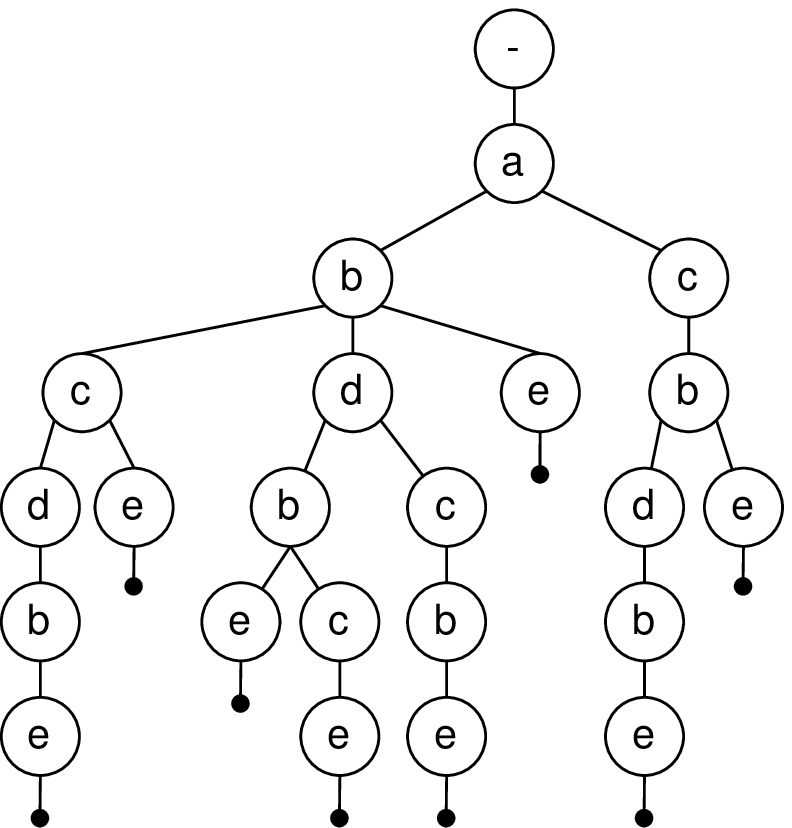}
    \caption{Running example: trie}
    \label{fig:runex_2_trie}
\end{figure}

A trie is a special type of tree, commonly referred to as a prefix tree, where all the children of a node have a common prefix. This paper utilizes the work done previously in \cite{awad2021efficient} for the construction of tries. For this purpose, the trie constructor needs a finite set of traces, i.e. a proxy log. For the running example, let us assume that we simulate a log from the process model in Figure~\ref{fig:petri:net}. If we allow for the arc that produced the loop to be traversed a single time, then we get maximally the following traces in our proxy log $L=\{
\langle a,b,c,d,b,e\rangle$, 
$\langle a,b,c,e\rangle$, 
$\langle a,b,d,b,e\rangle$, 
$\langle a,b,d,b,c,e\rangle$, 
$\langle a,b,d,c,b,e\rangle$, 
$\langle a,b,e\rangle$, 
$\langle a,c,b,d,b,e\rangle$, 
$\langle a,c,b,e\rangle$\}.

The proxy log represents all behavior allowed by the model, provided that no more than a single loop traversal is done. Using this proxy log, we can now construct a trie such as in Figure~\ref{fig:runex_2_trie}.

\begin{definition}[Proxy Trie]\label{def:trie} 
Let 
$M'\subseteq M$ be some proxy (sample) behavior of a process model. Then, the
\emph{proxy trie} constructed for it is a structure 
$T = (N,E,root,l,isEnd,min,max)$ where:
\begin{itemize}
    
	\item $N$ is a finite set of nodes. There is one node per prefix ${}_i\pi$ 
	for any execution sequence $\pi \in M'$ as well as one additional node 
	$root\in N$.
	\item $E \subset N \times N$ is a set of edges, s.t. for all $n\in 
	N$ it holds $|\{n'\mid (n',n)\in E \}|\leq 1$ and $(N,E)$ is a connected 
	graph. There are edges from $root$ to all nodes representing prefixes 
	of length one, and from each node $n$ to node $n'$, if the prefix 
	represented by $n'$ is obtained from the 
	prefix of $n$ by concatenation with a single activity.
	\item $root \in N$ is the root of the trie, i.e., the only node $n\in N$ 
	for which $|\{n'\mid (n',n)\in E \}|=0$;
	\item $l: N \rightarrow (\mathcal{A} \cup \{\bot\})$ is a labeling function for nodes. 
 The label is the activity of the prefix represented by the node, while $root$ is assigned $\bot$.  
	\item $isEnd: N \rightarrow \{0,1\}$ is a Boolean function that indicates 
	end nodes, i.e., whether the prefix represented by the node denotes an 
	execution sequence. 
	\item $min, max: E \rightarrow \mathcal{N}$ are functions assigning  
	to an edge the minimum and maximum path length, respectively, 
	to reach an end node, when traversing that edge. 
\end{itemize}
\end{definition}

\subsection{Conformance Checking}
\label{sec:conformance:checking}
Conformance checking compares the behavior recorded in an event log $L$ with the 
behavior specified by a process model~\cite{DBLP:books/sp/CarmonaDSW18}, $M$. To 
this end, an alignment between a trace $\sigma \in L$ of the log and an 
execution sequence $\pi \in M$ the model may be 
computed~\cite{DBLP:conf/edoc/AdriansyahDA11}. An alignment 
$\gamma = \langle (x_1,y_1),\ldots,(x_n,y_n) \rangle $ is a sequence of steps, 
each step $(x,y)\in (\mathcal{A}\cup \{\gg\}) \times (\mathcal{A}\cup 
\{\gg\})$ linking an activity of the trace, or the skip symbol $\gg$, to an 
activity of the execution sequence, or the skip symbol, whereas a step 
$(\gg,\gg)$ is illegal. Here, it must hold that 
the projection of $\gamma$ on the first component, ignoring $\gg$, yields 
$\sigma$, and the projection of $\gamma$ on the second component, ignoring 
$\gg$, yields $\pi$. A step $(x,y)$ is called \emph{synchronous move} if 
$x,y\in\mathcal{A}$, \emph{log move} if $y=\gg$, 
a \emph{model move} if $x=\gg$, while the last two are jointly referred to as 
asynchronous moves. 

Assigning costs to steps, a cost-optimal alignment of a trace and 
an execution sequence can be identified. Often, a cost of one is assigned 
to asynchronous moves, while synchronous moves for the same 
activity have zero cost. Moves on silent $\tau$ transitions can never be observed in the trace, and are thus also assigned a cost of zero. 
An optimal alignment minimizes the edit distance 
between a trace and an execution sequence. Identifying an optimal alignment for 
a trace and all execution sequences of a model is computationally expensive~\cite{DBLP:books/sp/CarmonaDSW18}. A synchronous product net is constructed from the combination of the process model and a model representation of the trace. It then becomes a search problem for finding the shortest path, commonly utilizing the A* algorithm and showing an exponential time complexity in the size of the trace and the model~\cite{van_der_aalst_replaying_2012}. 
In the remainder, we write $\delta(\gamma)$ for the total cost of an alignment~$\gamma$.


For our running example, let us assume that we have observed a trace $\sigma=\langle a,b,b,c\rangle$. The alignment between this trace and the trie is shown in Table \ref{tab:alignment_example}. Each activity in the trace and the model are paired in what is called a \emph{move}. The top row shows the complete \emph{trace} that has been seen. The bottom row shows the corresponding moves in the \emph{model}. In this case, it can be seen that the second execution of $b$ was not allowed by the model, and thus a log move was made. After executing a synchronous move on $c$, a model move on $e$ needs to be executed as well in order for the execution sequence to conclude. 
An alignment can also be \emph{suboptimal}, in case the cost associated with an alignment is not the minimal number of edit operations required to align the trace and the model. In this paper, an alignment refers to an optimal alignment, unless otherwise stated. Approximate algorithms commonly produce alignments that may not be optimal. Thus, such algorithms introduce an \emph{error} and their respective alignment costs can be compared, with a lower alignment cost indicating a closer result to an \emph{optimal} alignment.

A known impediment of alignments is that there can be several optimal alignments with equal cost but different operations. This can lead to nondeterministic decisions in terms of interpreting the alignment. However, for the purposes of this paper, handling this impediment is considered out of scope.

\begin{table}[]
\centering
\begin{tabular}{cl|c|c|c|c|c|}
\cline{3-7}
Trace &  & a & b & b  & c & $\gg$ \\ \cline{3-7} 
Model &  & a & b & $\gg$ & c & e  \\ \cline{3-7} 
\end{tabular}
\caption{An alignment between a trace and a model.}
\label{tab:alignment_example}
\end{table}

\begin{table}[]
\centering
\begin{tabular}{clccccl}
\cline{3-6}
Trace                & \multicolumn{1}{l|}{\textbf{}} & \multicolumn{1}{c|}{a} & \multicolumn{1}{c|}{b} & \multicolumn{1}{c|}{b}                  & \multicolumn{1}{c|}{c} &                        \\ \cline{3-6}
Model                & \multicolumn{1}{l|}{}          & \multicolumn{1}{c|}{a} & \multicolumn{1}{c|}{b} & \multicolumn{1}{c|}{$\gg$} & \multicolumn{1}{c|}{c} &                        \\ \cline{3-6}
\multicolumn{1}{l}{} &                                & \multicolumn{1}{l}{}   & \multicolumn{1}{l}{}   & \multicolumn{1}{l}{}                    & \multicolumn{1}{l}{}   &                        \\ \cline{3-7} 
Trace                & \multicolumn{1}{l|}{\textbf{}} & \multicolumn{1}{c|}{a} & \multicolumn{1}{c|}{b} & \multicolumn{1}{c|}{$\gg$} & \multicolumn{1}{c|}{b} & \multicolumn{1}{l|}{c} \\ \cline{3-7} 
Model                & \multicolumn{1}{l|}{}          & \multicolumn{1}{c|}{a} & \multicolumn{1}{c|}{b} & \multicolumn{1}{c|}{d}                  & \multicolumn{1}{c|}{b} & \multicolumn{1}{l|}{c} \\ \cline{3-7} 
\end{tabular}
\caption{Two optimal prefix-alignments for the running example.}
\label{tab:prefix-alignment_example}
\end{table}



\subsubsection{Prefix-alignment}

In an online setting, the trace execution may not yet have concluded and it is unknown how the execution sequence might play out. In such cases, an alignment - i.e., a \emph{complete} alignment - overestimates the actual conformance cost. A prefix-alignment is a variation of the alignment where complete path traversal to the model's sink is not necessary. Returning to the trace $\sigma=\langle a,b,b,c\rangle$ one can deduce that the final event $e$ may still occur, and thus there exists no deviation in terms of the activity $e$. In this case, there exist two equally optimal prefix-alignments. Either an assumption can be made that the second $b$ should not have happened, or the third activity $d$ was not conducted. Prefix-alignments $\hat{\gamma}$ of trace $\sigma$ are shown in Table~\ref{tab:prefix-alignment_example}.





\section{Related works} \label{sec:related}
For a more thorough background on conformance checking, we refer to \cite{carmona2018conformance}. Much of the research in conformance checking of the past decade has relied on the concept of alignments \cite{van_der_aalst_replaying_2012}. Alignments are a mapping between the moves in the log (actual behavior) and possible moves in the model. Generally, alignments are considered to provide good diagnostics, as it is easy to interpret common behavior and deviations, such as skipping of an activity or conducting an activity where it was not expected by the model~\cite{van2022process}. While alignments generally allow for a concise understanding of the conformance, the calculation of an alignment is time-consuming, as the most common way of calculating an alignment is by building a so-called synchronous product net and using the A* search algorithm to find the shortest path through the net~\cite{van_der_aalst_replaying_2012}. When building the synchronous product net, the search space may grow exponentially. Thus, the calculation of \emph{optimal} alignments is generally not considered practical in real-life settings.

\subsection{Approximation methods}
In order to allow for acceptable execution times, approximation methods have been introduced. Generally, utilizing an approximation method results in a small loss in accuracy, but a big gain in computation time. 

Sampling an event log has been studied in an effort to efficiently calculate the approximate conformance of the whole event log~\cite{kabierski2021sampling}. The main idea is to find non-conforming behavior from an event log by utilizing guided sampling. The proposed methods are efficient for finding the conformance of the event log but are not directly applicable to online conformance checking, where a specific deviation would need to be pinpointed.

In~\cite{fani2020conformance}, the authors used methods for constructing a finite sample of traces achievable by the process model, i.e., a sample of the process model, the so-called proxy model. With the finite proxy model and the by-nature finite log, Levenshtein string edit distance can be used to find the best approximate alignment. This is done by iterating over the traces in the log and comparing against each trace in the proxy model and reporting the model trace with the smallest edit distance. While the method is suitable for finding specific deviations, the edit distance computation time for larger models and traces renders this method unfeasible for online scenarios.

An approximate algorithm using the trie data structure was introduced in~\cite{awad2021efficient}. The trie data structure is especially efficient in reducing the search space and the approximate algorithm is, in some cases, able to compute alignments in an offline setting several magnitudes faster than the edit distance, while only adding a moderate error. While the computation time of this method would make it feasible for online conformance checking, the approximate algorithm has been purposed for static event logs where the whole trace is known beforehand.

Notably, all of the approximation methods thus far have been applied for calculating conformance in an offline setting. As indicated in Section~\ref{sec:introduction}, offline conformance checking deals with past data and can thus suffer from incorrect penalization of ongoing cases. More importantly, the value of offline conformance diminishes with time, as many business processes expect prompt responses to deviations. Thus, relevant online conformance checking methods are discussed next.


\subsection{Online Conformance Checking}

The general framework for online conformance checking was introduced in~\cite{burattin2017framework}. The work introduced an approach for online conformance checking, generally recognized as the first approach that is able to calculate conformance in near real-time. Since then, two research directions have evolved that have investigated online conformance checking methods and artifacts of online conformance checking.

The work in~\cite{burattin2018online} used behavioral patterns for calculating conformance in an online setting. Most notably, in addition to conformance, the method outputs completeness and confidence metrics. These metrics give additional insights to the user in terms of the reliability of the conformance. Also, the behavioral methods do not penalize \emph{warm starting} scenarios. That is cases where a process execution has been started before the conformance checking begins. More recently, ~\cite{lee2021orientation} extended the behavioral approach, basing their method on Hidden Markov Models, alternating between state estimations and calculating conformance. While the methods in this direction are very fast in terms of computation time, they are less informative in terms of diagnosing the causes of deviations. Generally, these methods can be considered as trace-level metrics, indicating whether something is wrong and how trustworthy the assessment is. It is hard to pinpoint what exactly is the non conforming part between the trace and the model.

The other main path has focused on prefix-alignments, which were first presented in~\cite{van2019online}. The algorithm included a window-size parameter as a way to trade-off between the computation time and cost optimality. An infinite window size allows for the calculation of optimal prefix-alignments but has the slowest execution time. A window size of one is the fastest, but the alignments produced may be suboptimal. In~\cite{schuster2020online}, the authors improved upon their work by introducing an incremental A* algorithm which is able to calculate optimal prefix-alignments with a smaller memory footprint. However, computationally the newer method remained noticeably slower than the initial algorithm with a window size of one. More recent work has seen proposals for various memory-efficient approaches for calculating prefix-alignments in an online setting~\cite{zaman2021efficient}. But in general, due to the reliance on computing synchronous product nets and then doing shortest path traversal, the prefix-alignment methods exhibit a heavy computation load and remain impractical for most real-life scenarios.

The work in this paper is a contribution to the latter research path, as the algorithm introduced here calculates prefix-alignments in an online setting. The target is to improve the computation time of prefix-alignments while not introducing a noticeable error, in order to increase the feasibility of using prefix-alignments in real-life settings.

\section{Approach} \label{sec:approach}

\begin{figure*}[h!]
    \centering
    \includegraphics[width=0.75\linewidth]{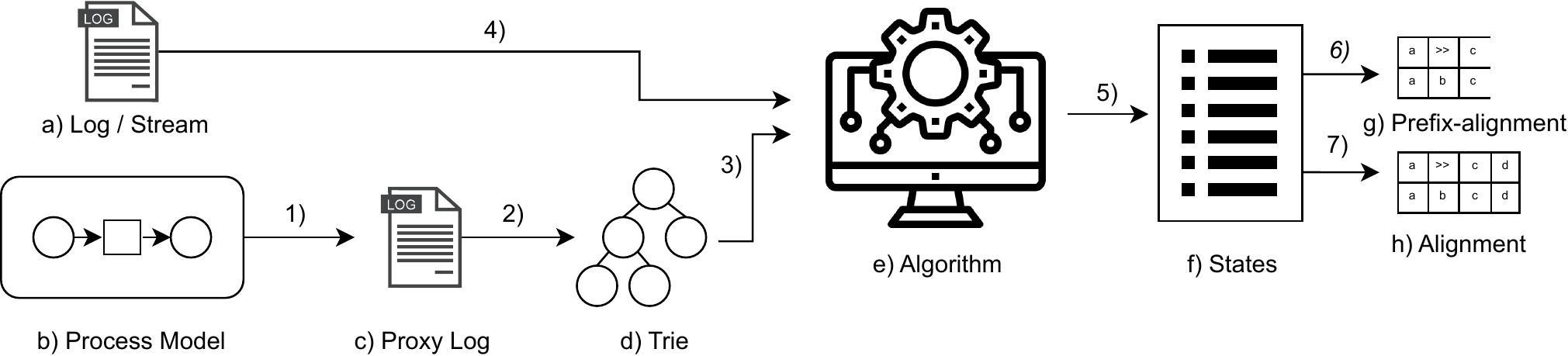}
    \caption{Approach overview}
    \label{fig:approach}
\end{figure*}

In this section, the approach and the algorithm are introduced. An overview of the approach is given in Figure \ref{fig:approach}.

To begin with, the approach assumes the existence of an event log or an event stream $(a)$ and a process model $(b)$. The process model is either sampled or simulated in step $(1)$ into a proxy log $(c)$. From the proxy log, step $(2)$ constructs the proxy trie $(d)$. The trie is given as input to the algorithm in step $(3)$ and then, the algorithm expects an event log or an event stream with case ID $c$ and a subsequence of events $\sigma^\prime$ in step $(4)$. The algorithm $(e)$ checks for conformance and in step $(5)$ outputs a list of States $(f)$. Finally, two optional steps step $(6)$ for fetching the latest Prefix-alignment $(g)$ for a case and step $(7)$ for calculating and fetching a complete alignment $(h)$ which is permissible by the list of States $(f)$. Importantly, as the algorithm holds a list of states with prefix-alignments, it is possible to use different methods, such as \cite{awad2021efficient} or \cite{fani2020conformance}, in step $(7)$ for finding the complete alignment from a prefix-alignment.

In an online setting, the events are expected to be processed one by one. Furthermore, it is common that multiple cases are ongoing at the same time - meaning that events coming in belong to different cases. The algorithm needs to be able to keep track of the seen cases and their states while performing optimizations for low memory consumption. For handling these demands, the definitions for a state, decay time, state buffer, and look-ahead limit are introduced.

\begin{definition}[State]\label{def:state}
A state $s$ is a tuple $(n, 
\hat{\gamma}, \hat{\sigma}, \delta(\hat{\gamma}), dt)$, where $n$ is the current node in the trie,
$\hat{\gamma}$ is the prefix-alignment up to this node, 
$\hat{\sigma}$ is the trace suffix, 
$\delta(\hat{\gamma})$ is the total cost of the current state,
and $dt$ is the associated decay time of the state.
\end{definition}
The state holds the information necessary for the algorithm to compute the conformance - namely, the current node $n$ in the trie, and unprocessed trace suffix $\hat{\sigma}$.
For the running example, let us assume that we have seen so far the trace ${\sigma}=\langle a,b,b,c\rangle$. The most recent optimal State $s$ would thus have the current node $n = c$ where the path from the root is $abc$, as this is the model path in the prefix-alignment $\hat{\gamma}$ displayed in Table~\ref{tab:prefix-alignment_example}. The suffix $\hat{\sigma}$ = $\emptyset$, since $\hat{\gamma}$ contains the latest seen event $c$ and no event currently remains to be processed. The total cost $\delta(\hat{\gamma})$ = $1$, assuming that any non-synchronous move has a cost of 1. The decay time $dt$ value is determined by a hyperparameter, as discussed next.

\begin{definition}[Decay time] \label{def:decaytime}

Let $\mathbb{N} =\{1,2,3,\ldots,\mathbb{n}\}$ where $\mathbb{n}$ is a large natural number. Then, \emph{decay time} $dt \in \mathbb{N}$. $s.dt$ is the \emph{decay time} associated with a particular \emph{state}. With every arrival of a new event within scope of state $s$, $s.dt = s.dt - 1$. Let $S$ be the set of states kept in memory. If $s.dt < 1$, then $S = S \setminus \{s\}$.
\end{definition}

We distinguish between two modes for initializing $dt$.

\emph{Fixed decay time} denotes a pre-determined integer for each new state. For example, all new states are initialized with $s.dt := 5$. This is effectively a window size parameter.

\emph{Discounted decay dime} relies on the presumption that deviations near the beginning of a trace are more costly than deviations near the end of a trace \cite{boltenhagen2021discounted}. The equation for calculating the discounted decay time is given in Equation \ref{equ:discdecaytime}. 

\begin{equation}\label{equ:discdecaytime}
Max(\lfloor(\overline{T_{leaf}}-i)*df\rfloor,min_{dt})
\end{equation}

The hyperparameters are the \emph{discounting factor} $df$ and a \emph{minimum decay time} $min_{dt}$.
The average length from the root of the trie to each of the leaf nodes is marked by $\overline{T_{leaf}}$, and the current length of the trace is indicated by $i$ as in $\sigma(i)$, where $i$ indicates the $i$-th event of $\sigma$.

The default values set for the algorithm in this paper are $df = 0.3$ and $min_{dt} = 3$. As an example, if $\overline{T_{leaf}} = 100$, then if $i = 1$, i.e. it is the first event of a trace, then $dt = Max(\lfloor(100-1)*0.3\rfloor,3)$ = $30$. If $i = 50$, then $dt = 15$. If $i > 86$, then $dt = 3$, as $dt$ will effectively remain at the value set for $min_{dt}$.


\begin{definition}[State buffer]\label{def:statesbuffer}
Let $C$ be the set of \emph{case IDs} the algorithm has seen, and let $S$ be the set of states associated with a \emph{case ID} $c \in C$, while $\mathcal{P}(S)$ is the powerset of all the sets of states. The \emph{state buffer} B is then a mapping $ B:C\mapsto \mathcal{P}(S)$.
\end{definition}

The \emph{State Buffer} is updated with the arrival of every new event $e$ for case $c$. The current states of the \emph{case ID} are appended with the new event. That is $\forall s\in B(c)~s.\hat{\sigma} = s.\hat{\sigma} \cup \{e\}$. From each State $s \in S$, the associated costs for adding $e$ are calculated. New states with the least cost are added to the \emph{state buffer}.

An example can be seen in Table~\ref{table:statesbuffer} where the events $a,b,b,c$ arrive for a case and the states are calculated based on the trie from Figure \ref{fig:runex_2_trie}. $a$ is the first event for this case, thus the root state with id $0$ is added to the \emph{state buffer}. $a$ is a child of the trie's root, so the state with a synchronous move $(a,a)$ is also added to the \emph{state buffer} with state id $1$.

\begin{table}[]
\centering

\scalebox{0.8}{
\begin{tabular}{@{}ccccccc@{}}
\toprule
\textbf{\begin{tabular}[c]{@{}c@{}}Arriving\\      event\end{tabular}} & \textbf{\begin{tabular}[c]{@{}c@{}}State\\      id\end{tabular}} & \textbf{n} & \textbf{$\hat{\gamma}$}                                         & \textbf{$\hat{\sigma}$}                 & \textbf{$\delta(\hat{\gamma})$} & \textbf{dt} \\ \midrule
\multirow{2}{*}{a}                                                     & 0                                                                & -          & -                                                    & $\langle a\rangle$   & 0        & 2  \\
                                                                       & 1                                                                & $a$          & (a,a)                                                & -                            & 0        & 2  \\ \midrule
\multirow{3}{*}{b}                                                     & 0                                                                & -          & -                                                    & $\langle a,b\rangle$ & 0        & 1  \\
                                                                       & 1                                                                & $a$          & (a,a)                                                & $\langle b\rangle$   & 0        & 1  \\
                                                                       & 2                                                                & $_ab$         & (a,a)(b,b)                                           & -                            & 0        & 2  \\ \midrule 
\multirow{3}{*}{b}                                                     & 2                                                                & $_ab$         & (a,a)(b,b)                                           & $\langle b\rangle$   & 0        & 1  \\
                                                                       & 3                                                                & $_ab$         & (a,a)(b,b)(b,$\gg$)           & -                            & 1        & 2  \\
                                                                       & 4                                                                & $_{abd}b$       & (a,a)(b,b)($\gg$,d)(b,b)      & -                            & 1        & 2  \\ \midrule
\multirow{4}{*}{c}                                                     & 3                                                                & $_ab$         & (a,a)(b,b)(b,$\gg$)           & $\langle c\rangle$   & 1        & 1  \\
                                                                       & 4                                                                & $_{abd}b$       & (a,a)(b,b)($\gg$,d)(b,b)      & $\langle c\rangle$   & 1        & 1  \\
                                                                       & 5                                                                & $_{ab}c$        & (a,a)(b,b)(b,$\gg$)(c,c)      & -                            & 1        & 2  \\
                                                                       & 6                                                                & $_{abd}b$      & (a,a)(b,b)($\gg$,d)(b,b)(c,c) & -                            & 1        & 2  \\ \bottomrule
\end{tabular}
}
\caption{Running example: state buffer. The prefix path of the node is written in subscript.}
\label{table:statesbuffer}
\end{table}


Even though only new states with the least cost are included, preserving a \emph{state buffer} puts strain on the computer memory, as with each new event arrival, we need to store at least one, but possibly many new states in the buffer. Thus, the \emph{Decay Time} is decremented with each new event arrival for the associated \emph{case ID}.

Lastly, as part of the algorithm, we introduce the \emph{Look-Ahead Limit} for speeding up calculation time in case model moves are needed. 

\begin{definition}[Look-Ahead Limit]\label{def:lookaheadlimit}
Let $|\hat{\sigma}|$ be the size of the trace suffix, and $s.n.level$ the level of the current state's node in the trie. Then, the \emph{Look-Ahead Limit} $lim = |\hat{\sigma}| + s.n.level + 1$.
\end{definition}

The \emph{Look-Ahead Limit} is used for handling model moves, which are more complex in an online setting as the algorithm has to assume the model move is at least as useful as making a log move. In order to limit a potentially costly traversal, a model move should be realized iff we get a full substring match to $\hat{\sigma}$ in the paths below $s.n$ such that the first matching node is at most at the level $lim$. 

\begin{figure}[h!]
    \centering
    \includegraphics[width=0.7\linewidth]{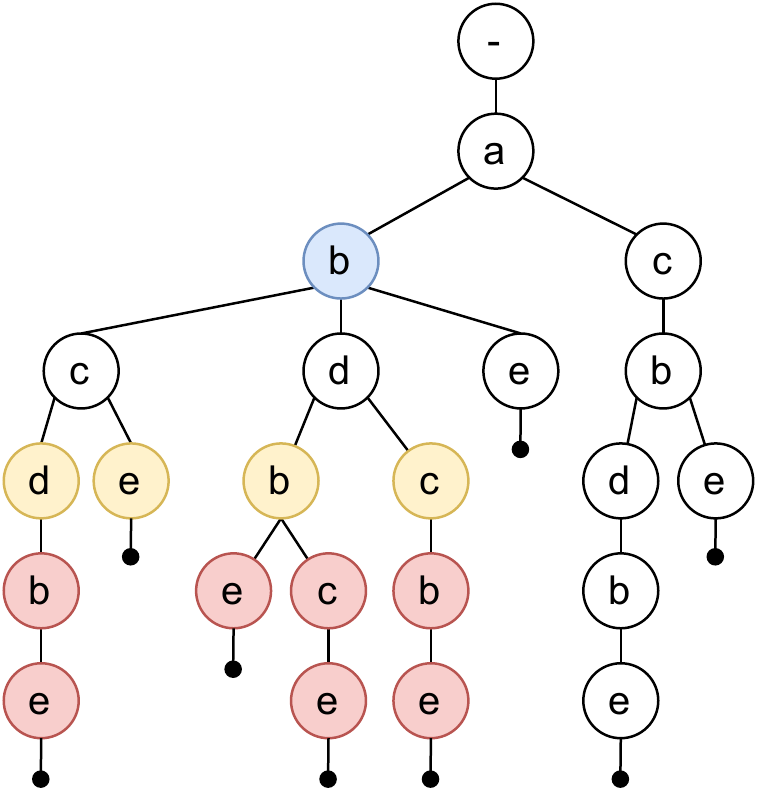}
    \caption{Running example: look ahead limit for state $id = 2$ when the second $b$ event is seen. The current node is shown in blue, the look-ahead limit in yellow, and nodes out of consideration in red.}
    \label{fig:look_ahead_limit}
\end{figure}

Based on the running example and states from Table \ref{table:statesbuffer}, when receiving the second $b$, the state $id = 2$ cannot utilize a synchronous move, as it is currently at node $n = b$ where $ab$ is the current path in the trie. $|\hat{\sigma}| = 1$, as this is the second $b$ that is not processed by state $id = 2$. $s.n.level = 2$, as the node is $2$ steps from the $root$ node. The look-ahead limit for state $id = 2$ is thus $lim = 1 + 2 + 1 = 4$. This indicates that the algorithm should attempt to make model moves iff it gets a substring match on event $b$ at most $4$ steps from the $root$. From the Figure~\ref{fig:look_ahead_limit} it can be deduced that the path $abdb$ is the only viable model move path to get a synchronous move on the second $b$.

\subsection{Implementation}

\begin{algorithm}
\caption{I Will Survive}
{\fontsize{9}{9}\selectfont
\begin{algorithmic}[1]
\Require $c, \sigma^\prime, T, B, df, min_{dt}$

\State $S \gets \emptyset$

\If{ $c \in B$} \label{line:checkbuffer:start}
    \State $S \gets B(c)$
\Else
    \State $s \gets (s.n = T.root,\newline
            \hspace*{5em} s.\hat{\gamma} = {\emptyset}\newline
            \hspace*{5em} s.\hat{\sigma} = {\emptyset},\newline
            \hspace*{5em} s.\delta(\hat{\gamma}) = 0,\newline
            \hspace*{5em} s.dt = f(\overline{T_{leaf}}, i=0, df, min_{dt})$ \Comment{Decay Time is calculated as in Equation \ref{equ:discdecaytime}}
    \State $S \gets \{s\}$ 
\EndIf \label{line:checkbuffer:end}

\ForEach {$e \in \sigma^\prime$} \Comment{Loop in case of multiple events}
    \State $S_{sync} \gets \emptyset$ \Comment{placeholder for sync moves}
    \State $S_{interim} \gets \emptyset$ \Comment{placeholder for log \& model moves}
    \ForEach {$s \in S$}
        \If{$s.syncPossible(e)$} \label{line:syncpossible} \Comment{Attempt to make a sync move on event e}
            \State $s_{new} \gets (s_{new}.n = s.n.getChild(e), \newline
            \hspace*{8em} s_{new}.\hat{\gamma} = s.\hat{\gamma}\cup \{(e,e)\}, \newline
            \hspace*{8em} s_{new}.\hat{\sigma} = \emptyset, \newline
            \hspace*{8em} s_{new}.\delta(\hat{\gamma}) = s.\delta(\hat{\gamma}), \newline
            \hspace*{8em} s_{new}.dt = f(\overline{T_{leaf}}, i, df, min_{dt}))$ 
            \State $S_{sync} \gets S_{sync} \cup \{s_{new}\}$
        \EndIf
    \EndFor
    \If {$\vert S_{sync}\vert$ $>$ $0$}
        \ForEach{$s \in S$} \label{line:syncsfound:start}
            \If {$s.dt < 1$}
                \State $S \gets S \setminus \{s\}$
            \Else
               \State $s.\hat{\sigma} \gets s.\hat{\sigma} \cup \{e\}$
                \State $s.dt \gets s.dt - 1$ \label{line:syncsfound:end}
            \EndIf
        \EndFor
        \ForEach {$s_{new} \in S_{sync}$} \label{line:syncsadd:start}
            \State $S \gets S \cup \{s_{new}\}$ \label{line:syncsadd:end}
        \EndFor 
    \Else 
        \State $\delta(\hat{\gamma})_{min} \gets \mathbb{N}_{max} $ \label{line:mincostvar} \Comment{Instantiate a minimum cost variable with a large natural number}
        \ForEach {$s \in S$}
            \State $s_{log} \gets handleLogMove(s,e)$ \Comment{New state with log moves on $s.\hat{\sigma} \cup \{e\}$} \label{line:logmove}
            \If {$s_{log}.\delta(\hat{\gamma}) <= \delta(\hat{\gamma})_{min}$} \label{line:logadd:start}
                \State $S_{Interim} \gets S_{Interim} \cup \{s_{log}\}$
                \State $\delta(\hat{\gamma})_{min} \gets s_{log}.\delta(\hat{\gamma})$ \label{line:logadd:end}
            \EndIf
            \State $S_{model} \gets handleModelMoves(s,e)$ \label{line:modelmoves}\Comment{A set of new states with model moves}
            
            \ForEach{$s_{model} \in S_{model}$} \label{line:modelmovesadd:start}
                \If {$s_{model}.\delta(\hat{\gamma}) <= \delta(\hat{\gamma})_{min}$}
                \State $S_{Interim} \gets S_{Interim} \cup \{s_{model}\}$
                \State $\delta(\hat{\gamma})_{min} \gets s_{model}.\delta(\hat{\gamma})$
                \EndIf \label{line:modelmovesadd:end}
            \EndFor
            \If {$s.dt < 1$} \label{line:statedecay:start}
                \State $S \gets S \setminus \{s\}$
            \Else
                \State $s.\hat{\sigma} \gets s.\hat{\sigma} \cup \{e\}$
                \State $s.dt \gets s.dt - 1$
            \EndIf \label{line:statedecay:end}
            
        \EndFor
        \ForEach{$s_{interim} \in S_{interim}$} \label{line:interimadd:start}
            \If {$s_{interim}.\delta(\hat{\gamma}) == \delta(\hat{\gamma})_{min}$}
                \State $S \gets S \cup s_{interim}$
            \EndIf
        \EndFor \label{line:interimadd:end}
    \EndIf
    
\EndFor

\State $B.S \gets S$ \label{line:statesbufferupdate}


\end{algorithmic}
}
\label{alg:iws}
\end{algorithm}

The pseudo-code for the IWS algorithm is listed in Algorithm~\ref{alg:iws}. The algorithm takes as input the case ID $c$, the list of received events $\sigma^\prime$, the trie $T$, the state buffer $B$, the \emph{discounting factor} $df$ and a \emph{minimum decay time} $min_{dt}$.

First, in lines~\ref{line:checkbuffer:start}-\ref{line:checkbuffer:end}, the algorithm checks if the case ID is already in the buffer. If yes, then the case ID’s current states are fetched. If the case ID is not in the buffer, then an empty state at the root of the trie is initialized. Using the running example from Table~\ref{table:statesbuffer}, this is the place where state $id = 0$ is generated.

Then, the events in the seen trace are looped over. There can be zero up to n events. All current states are iterated to see if the activity is a child of any of the states' nodes (line~\ref{line:syncpossible}). In the running example, this corresponds to $Arriving event = a$ and state $id = 1$ being created. 
If synchronous moves are found, then all previous current states are updated by adding the activity to the suffix and decrementing the decay time of the state (lines~\ref{line:syncsfound:start}-\ref{line:syncsfound:end}). Then, newly found synchronous moves are added to the list of current states (lines~\ref{line:syncsadd:start}-\ref{line:syncsadd:end}).

If synchronous moves are not found, then log and model moves are attempted on all the current states. As an example, in Table~\ref{table:statesbuffer} the second arrival of event $b$ does not get a synchronous move on state $id = 2$. Thus, first, a minimum cost variable is instantiated with a large initial value (line~\ref{line:mincostvar}). Then, for handling log moves, the current activity is added to the state suffix, and the suffix is added as log moves to a new state (line~\ref{line:logmove}). This corresponds to the newly generated state $id = 3$ in the running example. If the new state with log moves has a cost that is equal to or lower than the current minimum cost variable, then it is added to the interim states and the minimum cost variable is updated (lines~\ref{line:logadd:start}-\ref{line:logadd:end}). 

For making model moves, the Look-Ahead Limit is utilized, until either full substring matches are found, or the Look-Ahead Limit is exhausted (line~\ref{line:modelmoves}). The algorithm for \emph{handleModelMoves} is shown in Algorithm~\ref{alg:modelmoves} and a running example is discussed in Section \ref{sec:approach}. The resulting states utilizing model moves are then added to the interim states, provided that their cost is less than or equal to the minimum cost variable (lines~\ref{line:modelmovesadd:start}-\ref{line:modelmovesadd:end}).
Then, the state's decay time is decremented, or if the decay time is less than one then it is dropped (lines~\ref{line:statedecay:start}-\ref{line:statedecay:end}).

After the loop over the current states is finished, the interim current states are looped, to make sure only the states with truly the lowest cost are added to the state buffer (lines~\ref{line:interimadd:start}-\ref{line:interimadd:end}). Finally, the case ID’s states in the state buffer are updated (line~\ref{line:statesbufferupdate}).

\begin{algorithm}
\caption{handleModelMoves}
{\fontsize{9}{9}\selectfont
\begin{algorithmic}[1]
\Require $s, e$

\State $\hat{\sigma}_{check} \gets s.\hat{\sigma}+e$ \label{line:vars:start}
\State $lim \gets |\hat{\sigma}_{check}| + s.n.level$
\State $N_{current} \gets {s.n}$
\State $N_{child} \gets \emptyset$
\State $N_{matching} \gets \emptyset$
\State $S_{matching} \gets \emptyset$ \label{line:vars:end}

\While{$lim > s.n.level$} \label{line:limit}
\ForEach{$n_{current} \in N_{current}$} \label{line:addchild:start}
\State $N_{child} \gets N_{child} \cup n_{current}.getChildren()$
\EndFor \label{line:addchild:end}
\ForEach{$n_{child} \in N_{child}$} \label{line:childsubstring:start}
\State $n_{matching} \gets n_{child}.getSubstringMatch(\hat{\sigma}_{check})$
\If{$n_{matching} \neq \emph{null}$}
\State $N_{matching} \gets N_{matching} \cup \{n_{matching}\}$
\EndIf
\EndFor \label{line:childsubstring:end}

\If{$|N_{matching}| > 0$}  \label{line:matchfound}
\BREAK
\EndIf

\State $N_{current} \gets N_{child}$  \label{line:varsupdate:start}
\State $N_{child} \gets \emptyset$
\State $lim \gets lim-1$  \label{line:varsupdate:end}

\If{$lim = 0 and |\hat{\sigma}_{check}| > 1$}  \label{line:splitsuffix:start}
\State $\hat{\sigma}_{check}.remove(0)$ \Comment{Remove the first element from the suffix}
\State $lim \gets |\hat{\sigma}_{check}| + s.n.level$           
\State $N_{current} \gets {s.n}$
\EndIf \label{line:splitsuffix:end}

\EndWhile

\If{$|N_{matching}| > 0$} \label{line:matchingnodes} \Comment{If there was no match, we return empty set}
\ForEach{$n_{matching} \in N_{matching}$} \label{line:constructmatching:start}
\State $\hat{\gamma}_{matching} \gets constructAlignment()$ \Comment{The alignment is constructed by finding synchronous moves, then model moves, and finally log moves.}
\State $\delta(\hat{\gamma})_{matching} \gets \hat{\gamma}_{matching}.calculateCost()$
\State $s_{matching} \gets (s_{matching}.n = n_{matching}, \newline
            \hspace*{8em} s_{matching}.\hat{\gamma} = \hat{\gamma}_{matching}, \newline
            \hspace*{8em} s_{matching}.\hat{\sigma} = \emptyset, \newline
            \hspace*{8em} s_{matching}.\delta(\hat{\gamma}) = \delta(\hat{\gamma})_{matching}, \newline
            \hspace*{8em} s_{matching}.dt = f(\overline{T_{leaf}}, i, df, min_{dt})$
\State $S_{matching} \gets S_{matching} \cup {s_{matching}}$
\EndFor
\EndIf \label{line:constructmatching:end}

\Return $S_{matching}$ \label{line:returnmatching}

\end{algorithmic}
}
\label{alg:modelmoves}
\end{algorithm}

Handling model moves is the most complex part of the algorithm. Making a synchronous move always returns the matching child with zero cost. Making a log move means that the suffix is processed while the state's node remains the same. However, when making model moves, there can be a multitude of paths which need to be traversed in the trie, and multiple states can have equal cost. To find these states, the algorithm needs to find \emph{matching nodes}, i.e. the nodes which allow for a substring match on the state's suffix.

The algorithm for handling model moves is shown in Algorithm~\ref{alg:modelmoves}. First, variables, including the Look-ahead Limit are initialized (lines~\ref{line:vars:start}-\ref{line:vars:end}). A while loop is initialized, which runs until the Look-ahead Limit allows (line~\ref{line:limit}). To find the set of matching nodes, first the children of the state's current node are found (lines~\ref{line:addchild:start}-\ref{line:addchild:end}) and then traversed to find the matching nodes (lines~\ref{line:childsubstring:start}-\ref{line:childsubstring:end}). If matching nodes are found, then the while loop is terminated (line~\ref{line:matchfound}). If no matching nodes are found, then the next level of nodes in the trie are queued in order to find matching nodes, and the Look-Ahead Limit is updated (lines~\ref{line:varsupdate:start}-\ref{line:varsupdate:end}). 

\begin{figure}[h!]
    \centering
    \includegraphics[width=0.2\linewidth]{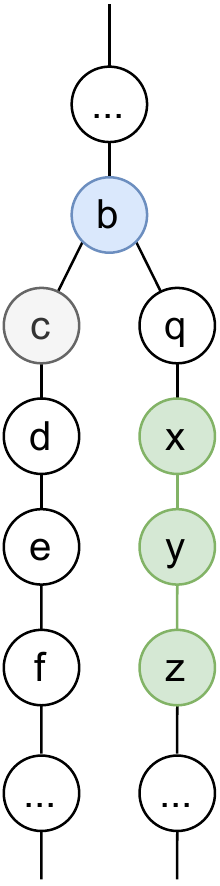}
    \caption{Example use case for look-ahead limit's suffix pruning. The last synchronous node is shown in blue, node $c$ is synchronous but leads to a suboptimal path, while nodes $x, y, z$ are more optimal, but do not get a substring match if starting from node $c$.} 
    \label{fig:look_ahead_suffix_split}
\end{figure}

If the Look-Ahead Limit is exhausted, but the suffix has more than 1 element, then the algorithm prunes the first element from the suffix and reinitalizes the finding of matching nodes (lines~\ref{line:splitsuffix:start}-\ref{line:splitsuffix:end}). An example for why this is necessary is illustrated in Figure~\ref{fig:look_ahead_suffix_split}. Arriving events $\langle b, c, x, y, z \rangle$ would not allow for a substring match starting from node $b$, because $\langle c, x, y, z \rangle$ is not a matching substring. However, removing $c$ the algorithm is able to find a match on $x, y, z$. Ultimately, this leads to the alignment $\langle(b,b), (c,\gg), (\gg,q), (x,x), (y,y), (z,z)\rangle$.

Once the while loop has concluded, the algorithm checks if matching nodes have been found (line~\ref{line:matchingnodes}). If matching nodes were found, then new matching states are constructed (lines~\ref{line:constructmatching:start}-\ref{line:constructmatching:end}). Ultimately, the algorithm returns the set of matching nodes (line~\ref{line:returnmatching}). In case no matching nodes were found, this is an empty set.

\subsection{Complexity analysis}

\subsubsection{Space complexity}
The two objects that need to be stored in memory are the trie and the state buffer. The trie is in the worst case linear to the size of the proxy log, $O(|M'|)$, indicating that each trace in the proxy log has a unique first activity. Commonly, the trie is logarithmic compared to the proxy log size $O(log |M'|)$. The trie is computed beforehand and it is immutable.


The size of the state buffer depends on two factors, the number of cases $|C|$ in the event stream, and the number of states stored for each case. The number of new states generated is dependent on the branching factor of the trie and the deviation in the behavior between a trace and the trie. In the best case, when an alignment consists of synchronous moves, the state buffer grows as $O(|C|.dt)$ that is because for each newly arriving event one new state is generated with a synchronous move. In the worst-case scenario, the states' growth can be equal to the branching factor, $bf$, of the trie node, i.e. $O(|C|.(bf+1).dt)$, we have only one possible log move but $bf$ model moves. Storage of previous states can be controlled by the decay time setting. Using a fixed decay time, there is a fixed upper bound on the number of states stored per case, that can be computed based on the precomputed trie. Using a discounted decay time, the upper bound is still dependent on the trie, while for each individual case the upper bound diminishes as the case evolves. 

\subsubsection{Time complexity}
For each newly arriving event we fetch the relevant states in $O(1)$. We retrieve in the worst case $O(|C|.(bf+1).dt)$ states, as discussed under space complexity. Synchronous moves and log moves can be done in $O(1)$. Handling model moves is dependent on the trie branching factor, and the look-ahead limit. The look-ahead limit $lim$ can be bounded by the decay time setting, e.g. the size of the trace suffix can never be longer than the decay time. That is we need at the worst case $O(bf.min(dt,lim))$ steps to define the new states to be added to the state buffer.

\section{Experiments} \label{sec:results}
\subsection{Motivation}

The experiments were conducted in order to answer the following two research questions:

RQ1: How does the algorithm compare against the current state-of-the-art in terms of finding prefix-alignments?

RQ2: How does the algorithm handle a stress test of fast-paced streams in terms of processing time and memory utilization?

All of the executions were done on a single thread using Windows 10 running a CPU @ 1.60GHz, Java 8 and heap size set to 8GB.

The implementation in Java and the execution results are available on GitHub\footnote{\url{https://github.com/DataSystemsGroupUT/ConformanceCheckingUsingTries/tree/streaming}}.

The preliminaries and experimental results for answering both of the research questions are discussed in the next subsections.

\subsection{Comparative Analysis}
\subsubsection{Preliminaries}

In the following, the naming convention from \cite{schuster2020online} will be used, namely the current state-of-the-art will be referred to as OCC-W1. The algorithm introduced in this paper will be referred to as IWS (I Will Survive). 
The target for the comparative analysis is to examine how the IWS algorithm fares in terms of alignment cost and computation time. In \cite{schuster2020online}, it was shown that OCC-W1 has generally a marginal cost error compared to optimal prefix-alignments. Thus, achieving costs comparable to OCC-W1 would indicate a low cost error. For calculating the computation time, only the time taken to process each event is taken into account. This is done to mimic an online scenario, where the loading of a model is done beforehand. 
For the experiments, both of the algorithms were executed in an \emph{offline} mode. That is, the event log was loaded from a file and fed to the algorithm event by event. This was done in order to have a fair comparison by avoiding discrepancies from networking or other external factors.

\subsubsection{Datasets}
For running the experiments, some well-known synthetic and real-life process logs were used. The synthetic process logs\footnote{\url{https://github.com/PADS-UPC/RL-align/tree/master/data/originals/M-models}} contained also a reference Petri net process model. 
The real-life process logs were BPI 2012\footnote{\url{https://doi.org/10.4121/uuid:3926db30-f712-4394-aebc-75976070e91f}}, BPI 
2017\footnote{\url{https://doi.org/10.4121/uuid:5f3067df-f10b-45da-b98b-86ae4c7a310b}}, and BPI 2020 Travel Permits\footnote{\url{https://doi.org/10.4121/uuid:52fb97d4-4588-43c9-9d04-3604d4613b51}}, which do not have an associated reference model. 

\begin{figure}
    \centering
    \includegraphics[height=0.4\columnwidth]{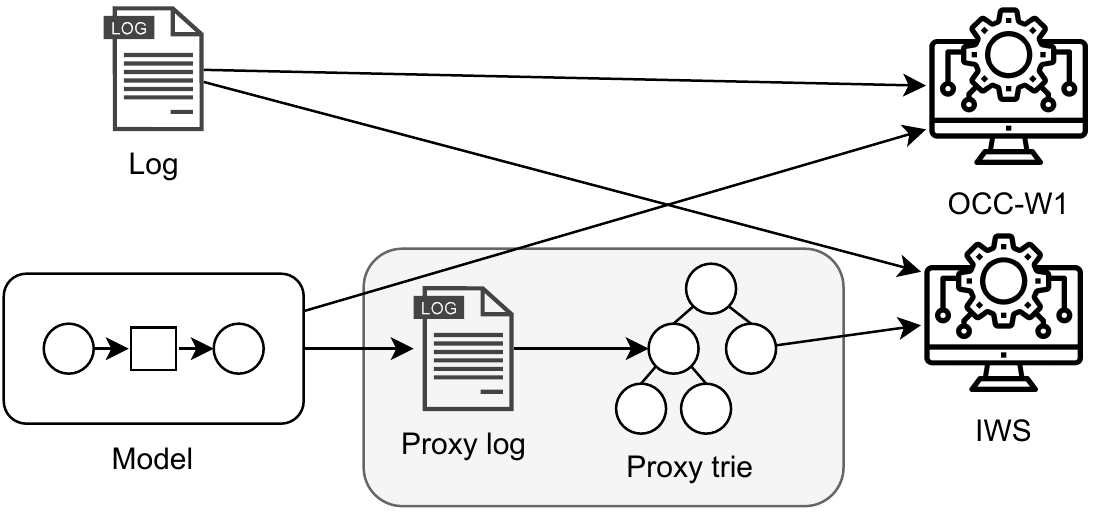}
	
    \caption{Setup for synthetic logs. Gray rectangle shows the artifacts produced by the preprocessing.}
    \label{fig:approach_simulated}
\end{figure}

The OCC-W1 takes as input an event and a petri net model, while IWS requires an event and a trie. Thus, some preprocessing was applied on both the synthetic and real-life datasets. 
The preprocessing for synthetic data is shown in Figure~\ref{fig:approach_simulated} with the gray rectangle indicating the additional steps done. For OCC-W1, the existing log and model were used. For IWS, a proxy log was simulated from the reference model. The simulation method from \cite{vanden2012improved} was used with default settings of random path simulation, 2000 generated traces and a maximum looping factor of 3. From the proxy log, the proxy trie was constructed and fed into the IWS algorithm, together with the original log.

\begin{figure}
    \centering
    \includegraphics[height=0.4\columnwidth]{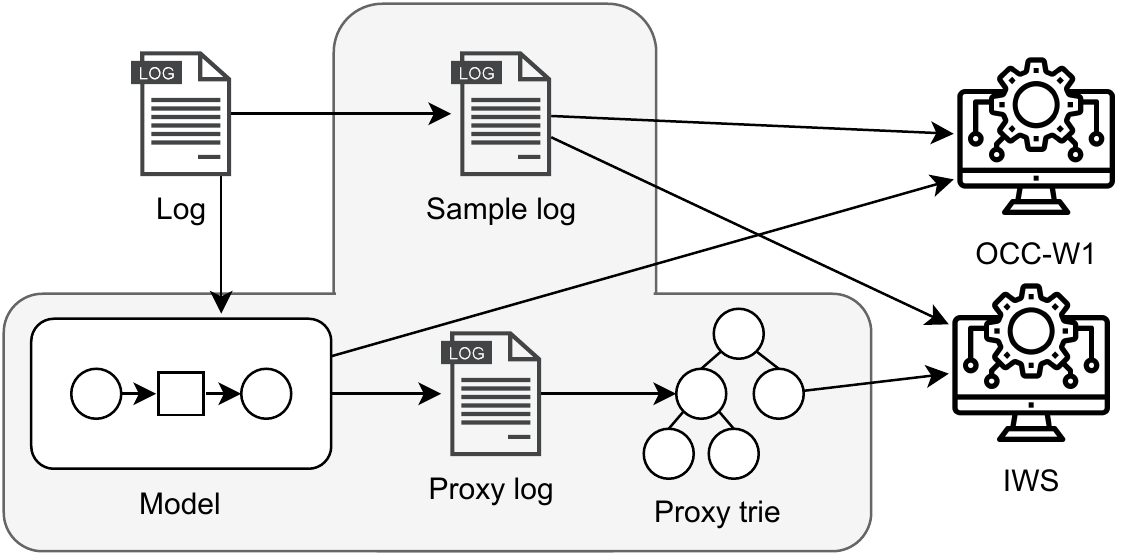}
	
    \caption{Setup for real-life logs. Gray rectangle shows the artifacts produced by the preprocessing.}
    \label{fig:approach_bpi}
\end{figure}

For the real-life data, the steps are shown in Figure~\ref{fig:approach_bpi}. The first step was to construct a process model from the log. For this, the Inductive Miner (IM) \cite{leemans2013discovering} plugin in ProM \cite{van2005prom} was used with noise thresholds set to 0.2, 0.5, 0.8 and 0.95. For the generation of the trie, the same steps with the same settings were done as for the synthetic data. Finally, while running the experiments it appeared that the OCC-W1 algorithm was unable to output a result due to the size of some of the original logs. Thus, sample logs of 1000 most frequent trace variants were generated and the algorithms used the sample logs instead of the original logs.

Information about the logs and models used in the experiments is shown in Table~\ref{table:logsizes} and Table~\ref{table:modelsizes}, respectively. Table~\ref{table:modelsizes} also shows information about the trie in terms of the amount of nodes and the construction time in milliseconds.

\begin{table}
\centering
\begin{tabular}{c|cc}
\textbf{Log} & \textbf{\# traces} & \textbf{\# events}  \\ 
\hline
M1           & 500                & 6555                \\
M2           & 500                & 8809                \\
M3           & 500                & 17980               \\
M4           & 500                & 13421               \\
M5           & 500                & 17028               \\
M6           & 500                & 26719               \\
M7           & 500                & 18803               \\
M8           & 500                & 8246                \\
M9           & 500                & 22163               \\
M10          & 500                & 29118               \\
BPI2012 sample      & 1000               & 33509               \\
BPI2017 sample      & 1000               & 32646               \\
BPI2020 sample      & 1000               & 15810              
\end{tabular}
\caption{Log metadata}
\label{table:logsizes}
\end{table}

\begin{table}
\centering
\caption{Model and trie metadata}
\label{table:modelsizes}
\begin{tabular}{c|c|c|c|c}
\textbf{Model} & \multicolumn{1}{c}{\textbf{Transitions}} & \begin{tabular}[c]{@{}c@{}}\textbf{Silent }\\\textbf{transitions}\end{tabular} & \multicolumn{1}{c}{\begin{tabular}[c]{@{}c@{}}\textbf{Trie}\\\textbf{nodes}\end{tabular}} & \begin{tabular}[c]{@{}c@{}}\textbf{Trie constr.}\\\textbf{time (ms)}\end{tabular}  \\
M1             & 39                                       & 3                                                                              & 8281                                                                                      & 353                                                                                \\
M2             & 34                                       & 2                                                                              & 20742                                                                                     & 407                                                                                \\
M3             & 123                                      & 14                                                                             & 26434                                                                                     & 524                                                                                \\
M4             & 52                                       & 8                                                                              & 13261                                                                                     & 512                                                                                \\
M5             & 33                                       & 1                                                                              & 49571                                                                                     & 652                                                                                \\
M6             & 72                                       & 2                                                                              & 98781                                                                                     & 1239                                                                               \\
M7             & 62                                       & 4                                                                              & 46481                                                                                     & 812                                                                                \\
M8             & 15                                       & 0                                                                              & 1682                                                                                      & 326                                                                                \\
M9             & 55                                       & 0                                                                              & 7223                                                                                      & 576                                                                                \\
M10            & 146                                      & 3                                                                              & 87289                                                                                     & 1197                                                                               \\
BPI 2012-0.2   & 46                                       & 22                                                                             & 3962                                                                                      & 1659                                                                               \\
BPI 2012-0.5   & 46                                       & 23                                                                             & 1721                                                                                      & 387                                                                                \\
BPI 2012-0.8   & 25                                       & 8                                                                              & 726                                                                                       & 262                                                                                \\
BPI 2012-0.95  & 24                                       & 4                                                                              & 106                                                                                       & 219                                                                                \\
BPI 2017-0.2   & 46                                       & 21                                                                             & 3338                                                                                      & 276                                                                                \\
BPI 2017-0.5   & 29                                       & 8                                                                              & 192                                                                                       & 150                                                                                \\
BPI 2017-0.8   & 32                                       & 7                                                                              & 69                                                                                        & 163                                                                                \\
BPI 2017-0.95  & 31                                       & 6                                                                              & 69                                                                                        & 152                                                                                \\
BPI 2020-0.2   & 85                                       & 37                                                                             & 23603                                                                                     & 397                                                                                \\
BPI 2020-0.5   & 64                                       & 15                                                                             & 945                                                                                       & 71                                                                                 \\
BPI 2020-0.8   & 65                                       & 16                                                                             & 887                                                                                       & 64                                                                                 \\
BPI 2020-0.95  & 59                                       & 14                                                                             & 1591                                                                                      & 235                                                                               
\end{tabular}
\caption{Model and trie metadata}
\label{table:modelsizes}
\end{table}

\subsubsection{Results}
The results of the experiments are shown in Table~\ref{table:comparativeresults}. For each of the datasets, the average alignment cost per trace is reported. The OCC-W1 is assumed to report a prefix-alignment cost close to the optimal. In terms of computation time, a time per trace and time per event in milliseconds is reported.

\begin{table*}
\centering
\scalebox{0.63}{
\begin{tabular}{|cc|ccc|ccc|cc|} 
\hline
\multirow{3}{*}{\textbf{Dataset}} & \multicolumn{1}{c}{\multirow{3}{*}{\begin{tabular}[c]{@{}c@{}}\textbf{IM }\\\textbf{threshold}\end{tabular}}} & \multicolumn{3}{c|}{\textbf{IWS}}                                                                                                                                                                                                                                                                 & \multicolumn{3}{c|}{\textbf{OCC-W1}}                                                                                                                                                                                                                                                              & \multicolumn{1}{c|}{\multirow{3}{*}{\begin{tabular}[c]{@{}c@{}}\textbf{Cost}\\\textbf{IWS/OCC-W1}\end{tabular}}} & \multicolumn{1}{c|}{\multirow{3}{*}{\begin{tabular}[c]{@{}c@{}}\textbf{Time}\\\textbf{IWS/OCC-W1}\end{tabular}}}  \\
                              & \multicolumn{1}{c}{}                                                                                          & \multirow{2}{*}{\begin{tabular}[c]{@{}c@{}}\textbf{Cost per }\\\textbf{trace}\end{tabular}} & \multirow{2}{*}{\begin{tabular}[c]{@{}c@{}}\textbf{Time per }\\\textbf{trace (ms)}\end{tabular}} & \multirow{2}{*}{\begin{tabular}[c]{@{}c@{}}\textbf{Time per }\\\textbf{event (ms)}\end{tabular}} & \multirow{2}{*}{\begin{tabular}[c]{@{}c@{}}\textbf{Cost per }\\\textbf{trace}\end{tabular}} & \multirow{2}{*}{\begin{tabular}[c]{@{}c@{}}\textbf{Time per }\\\textbf{trace (ms)}\end{tabular}} & \multirow{2}{*}{\begin{tabular}[c]{@{}c@{}}\textbf{Time per }\\\textbf{event (ms)}\end{tabular}} & \multicolumn{1}{c|}{}                                                                                            & \multicolumn{1}{c|}{}                                                                                             \\
                              & \multicolumn{1}{c}{}                                                                                          &                                                                                             &                                                                                                  &                                                                                                  &                                                                                             &                                                                                                  &                                                                                                  & \multicolumn{1}{c|}{}                                                                                            & \multicolumn{1}{c|}{}                                                                                             \\ 
\hhline{|~~--------|}
M1                            &                                                                                                              & 5.8                                                                                         & 2.3                                                                                              & 0.2                                                                                              & 5.1                                                                                         & 19.5                                                                                             & 1.5                                                                                              & {\cellcolor[rgb]{0.361,0.678,0.361}}1.1                                                                          & {\cellcolor[rgb]{0.337,0.667,0.337}}0.12                                                                         \\
M2                            &                                                                                                              & 10.6                                                                                        & 3.1                                                                                              & 0.2                                                                                              & 8.8                                                                                         & 104.6                                                                                            & 5.9                                                                                              & {\cellcolor[rgb]{0.49,0.741,0.49}}1.2                                                                            & {\cellcolor[rgb]{0.267,0.631,0.267}}0.03                                                                         \\
M3                            &                                                                                                              & 23.9                                                                                        & 16.5                                                                                             & 0.5                                                                                              & -                                                                                           & -                                                                                                & -                                                                                                & -                                                                                                                & {\cellcolor[rgb]{0.267,0.631,0.267}}-                                                                            \\
M4                            &                                                                                                              & 22.1                                                                                        & 17.3                                                                                             & 0.6                                                                                              & 22.4                                                                                        & 393.6                                                                                            & 14.7                                                                                             & {\cellcolor[rgb]{0.361,0.678,0.361}}1.0                                                                          & {\cellcolor[rgb]{0.267,0.631,0.267}}0.04                                                                         \\
M5                            &                                                                                                              & 26.0                                                                                        & 23.8                                                                                             & 0.7                                                                                              & -                                                                                           & -                                                                                                & -                                                                                                & -                                                                                                                & {\cellcolor[rgb]{0.267,0.631,0.267}}-                                                                            \\
M6                            &                                                                                                              & 45.9                                                                                        & 632.5                                                                                            & 11.8                                                                                             & -                                                                                           & -                                                                                                & -                                                                                                & -                                                                                                                & {\cellcolor[rgb]{0.267,0.631,0.267}}-                                                                            \\
M7                            &                                                                                                              & 29.2                                                                                        & 24.5                                                                                             & 0.7                                                                                              & -                                                                                           & -                                                                                                & -                                                                                                & -                                                                                                                & {\cellcolor[rgb]{0.267,0.631,0.267}}-                                                                            \\
M8                            &                                                                                                              & 7.6                                                                                         & 3.0                                                                                              & 0.2                                                                                              & 7.2                                                                                         & 10.5                                                                                             & 0.6                                                                                              & {\cellcolor[rgb]{0.361,0.678,0.361}}1.1                                                                          & {\cellcolor[rgb]{0.471,0.737,0.471}}0.28                                                                         \\
M9                            &                                                                                                              & 29.2                                                                                        & 32.0                                                                                             & 0.7                                                                                              & 23.0                                                                                        & 974.9                                                                                            & 22.0                                                                                             & {\cellcolor[rgb]{0.49,0.741,0.49}}1.3                                                                            & {\cellcolor[rgb]{0.267,0.631,0.267}}0.03                                                                         \\
M10                           &                                                                                                              & 50.0                                                                                        & 485.1                                                                                            & 8.3                                                                                              & -                                                                                           & -                                                                                                & -                                                                                                & -                                                                                                                & {\cellcolor[rgb]{0.267,0.631,0.267}}-                                                                            \\
BPI 2012                      & 0.2                                                                                                           & 27.1                                                                                        & 16.8                                                                                             & 0.5                                                                                              & 0.3                                                                                         & 141.0                                                                                            & 4.2                                                                                              & {\cellcolor[rgb]{1,0.455,0.455}}81.3                                                                             & {\cellcolor[rgb]{0.337,0.667,0.337}}0.12                                                                         \\
BPI 2012                      & 0.5                                                                                                           & 26.6                                                                                        & 15.6                                                                                             & 0.5                                                                                              & 3.3                                                                                         & 256.3                                                                                            & 7.6                                                                                              & {\cellcolor[rgb]{1,0.773,0.361}}8.0                                                                              & {\cellcolor[rgb]{0.267,0.631,0.267}}0.06                                                                         \\
BPI 2012                      & 0.8                                                                                                           & 28.3                                                                                        & 12.7                                                                                             & 0.4                                                                                              & 26.1                                                                                        & 88.8                                                                                             & 2.7                                                                                              & {\cellcolor[rgb]{0.361,0.678,0.361}}1.1                                                                          & {\cellcolor[rgb]{0.337,0.667,0.337}}0.14                                                                         \\
BPI 2012                      & 0.95                                                                                                          & 30.1                                                                                        & 12.1                                                                                             & 0.4                                                                                              & 30.1                                                                                        & 216.7                                                                                            & 6.5                                                                                              & {\cellcolor[rgb]{0.361,0.678,0.361}}1.0                                                                          & {\cellcolor[rgb]{0.267,0.631,0.267}}0.06                                                                         \\
BPI 2017                      & 0.2                                                                                                           & 26.1                                                                                        & 14.5                                                                                             & 0.4                                                                                              & 1.7                                                                                         & 108.1                                                                                            & 3.3                                                                                              & {\cellcolor[rgb]{1,0.718,0.2}}15.3                                                                               & {\cellcolor[rgb]{0.337,0.667,0.337}}0.13                                                                         \\
BPI 2017                      & 0.5                                                                                                           & 25.3                                                                                        & 10.6                                                                                             & 0.3                                                                                              & 25.4                                                                                        & 142.9                                                                                            & 4.4                                                                                              & {\cellcolor[rgb]{0.361,0.678,0.361}}1.0                                                                          & {\cellcolor[rgb]{0.267,0.631,0.267}}0.07                                                                         \\
BPI 2017                      & 0.8                                                                                                           & 28.6                                                                                        & 10.4                                                                                             & 0.3                                                                                              & 28.8                                                                                        & 112.8                                                                                            & 3.5                                                                                              & {\cellcolor[rgb]{0.361,0.678,0.361}}1.0                                                                          & {\cellcolor[rgb]{0.267,0.631,0.267}}0.09                                                                         \\
BPI 2017                      & 0.95                                                                                                          & 28.6                                                                                        & 11.0                                                                                             & 0.3                                                                                              & 28.8                                                                                        & 93.5                                                                                             & 2.9                                                                                              & {\cellcolor[rgb]{0.361,0.678,0.361}}1.0                                                                          & {\cellcolor[rgb]{0.337,0.667,0.337}}0.12                                                                         \\
BPI 2020                      & 0.2                                                                                                           & 11.7                                                                                        & 4.9                                                                                              & 0.3                                                                                              & 6.4                                                                                         & 59.8                                                                                             & 3.8                                                                                              & {\cellcolor[rgb]{0.796,0.678,0.361}}1.8                                                                          & {\cellcolor[rgb]{0.267,0.631,0.267}}0.08                                                                         \\
BPI 2020                      & 0.5                                                                                                           & 10.7                                                                                        & 2.3                                                                                              & 0.1                                                                                              & 8.7                                                                                         & 20.8                                                                                             & 1.3                                                                                              & {\cellcolor[rgb]{0.49,0.741,0.49}}1.2                                                                            & {\cellcolor[rgb]{0.337,0.667,0.337}}0.11                                                                         \\
BPI 2020                      & 0.8                                                                                                           & 10.3                                                                                        & 3.3                                                                                              & 0.2                                                                                              & 11.9                                                                                        & 23.3                                                                                             & 1.5                                                                                              & {\cellcolor[rgb]{0.361,0.678,0.361}}0.9                                                                          & {\cellcolor[rgb]{0.337,0.667,0.337}}0.14                                                                         \\
BPI 2020                      & 0.95                                                                                                          & 12.1                                                                                        & 4.0                                                                                              & 0.3                                                                                              & 7.5                                                                                         & 17.9                                                                                             & 1.1                                                                                              & {\cellcolor[rgb]{0.796,0.678,0.361}}1.6                                                                          & {\cellcolor[rgb]{0.424,0.686,0.424}}0.23                                                                         \\
\hline
\end{tabular}
}
\caption{Comparative analysis results. Difference is shown as a division of IWS by OCC-W1. A difference below 1 indicates that IWS outperformed OCC-W1.}
\label{table:comparativeresults}
\end{table*}

For five of the synthetic logs, OCC-W1 was left running for 1 hour but no output was produced. These are marked with a \emph{-} in the table. For other synthetic logs, the cost deviations were modest. The highest cost deviation was reported for the M9 log, where OCC-W1 reported an average cost of 23.0 per trace, while IWS reported 29.2. This indicates a cost deviation of about 1.3x. For the M4 synthetic log, IWS outperformed OCC-W1 in terms of cost. This can be explained by the fact that OCC-W1 has a window size of 1, meaning that it tries to find the local optimum cost. IWS, utilizing a discounted Decay Time, is able to backtrack to a certain degree and, in some cases, find more optimal alignments than OCC-W1.

In terms of execution time, the results are strongly in favour of IWS. In almost all cases, IWS was able to process events in less than 1 millisecond, while OCC-W1 was able to process an event in less than a millisecond only for the dataset M8. The M8 dataset also exhibited the smallest difference in terms of execution time, as IWS finished execution in roughly 0.28x compared to OCC-W1. For M2, M4 and M9 datasets, IWS computed the result in only a fraction of time, compared to the OCC-W1 execution time. This is especially interesting in terms of, for example, dataset M4, where IWS outperformed OCC-W1 by both - producing alignments which were more optimal, and finding these alignments more than 22x faster. Finally, for the datasets M3, M5 and M7, IWS computed alignments in 8-13 seconds, while OCC-W1 was unable to output a result within 1 hour. This indicates an execution time difference of roughly 3 orders of magnitude.

For the real-life datasets, the results were a bit more varied in terms of cost. IWS displayed an especially poor cost performance for the BPI 2012 dataset where the model was discovered with IM threshold setting 0.2. Here, OCC-W1 reported a cost 81.3x smaller than IWS, which would indicate an unacceptable cost error. The reason for such poor performance is discussed in the next section. Some other poor results were for BPI 2012 with IM 0.5, and BPI 2017 with IM 0.2. However, for more than half of the datasets the cost error was comparable between IWS and OCC-W1. In fact, for 4 datasets, IWS again outperformed OCC-W1 by producing alignments which are more optimal.

In terms of execution time, IWS is faster across all datasets. The smallest difference is for BPI 2020 with IM setting 0.95, where IWS finished in 0.23x time taken by OCC-W1. The biggest execution time difference was for BPI 2012 with IM setting 0.95. Interestingly, for this dataset, both algorithms had the same alignments cost, but IWS was able to compute the alignments 17.9x faster than OCC-W1.

\subsubsection{Discussion}

The results indicate that IWS is well-suited for online conformance checking, beating the state-of-the-art in computation time, and in some cases also in terms of finding the optimal prefix-alignments. However, there were a few cases in the real-life datasets where the cost error of IWS was very high compared to OCC-W1. Let us investigate the reason for this with some examples from the BPI 2012 log.

First of all, as mentioned earlier, the Inductive Miner has a tendency to discover \emph{flower models}, i.e. models allowing any kind of behavior, if the noise threshold is set to a low level. In such cases, the allowed behavior in the model is at a very high level. IWS is dependent on the existence of a trie, which in turn is dependent on the existence of a proxy log - a set of behavior \emph{extracted} from the model. The more behavior the model allows for, the more difficult it is to extract a representative proxy log. 
One way to have a more representative proxy log would be to increase the size of the proxy log. However, for \emph{flower models}, this may be infeasible. The petri net model produced by IM for the BPI 2012 log with 0.2 setting has 46 transitions, out of which 24 transitions are labeled and 22 are silent. The first 2 labeled transitions are fixed, but after that, due to the silent transitions, almost any of the 22 labeled transitions can occur. Due to loops, the following transition can also be any of the 22 labeled transitions, including the label itself. Thus, with each new event, the possible behavior increases exponentially. For a trace with 10 events, assuming the first 2 events are always sequential, there can be $2+22^8=54875873538$ possible variants. As can be deduced from Table~\ref{table:logsizes}, the BPI 2012 log has roughly 33 events per trace. A simulation method to calculate a proxy log which would not find deviations becomes impractical from a computational point-of-view. Furthermore, it can well be argued that exercising conformance checking on a process model that allows any kind of behavior has no real value, since any kind of behavior is conforming.

\begin{table*}
\centering
\scalebox{0.55}{
\setlength\tabcolsep{1.5pt}
\begin{tabular}{cc|c|c|c|c|c|c|c|c|c|c|c|c|c|c|c|c|c|c|cccccccccccccccccccccccccccc} 
\cline{3-21}
\multirow{2}{*}{IWS
  alignment}    &                      & A                    & B                    & P                    & P                    & D                    & E                    & P                    & E                    & L                    & K                    & O                    & G                    & H                    & I                    & E                    & I                    & M                    & F                    & I                        &                        &                        &                          &                          &                        &                        &                        &                          &                          &                          &                        &                          &                          &                          &                          &                          &                          &                          &                        &                          &                          &                          &                        &                          &                          &                         &    \\ 
\cline{3-21}
                                    &                      & A                    & B                    & P                    & P                    & $\gg$                   & E                    & P                    & $\gg$                   & $\gg$                   & $\gg$                   & $\gg$                   & $\gg$                   & $\gg$                   & $\gg$                   & $\gg$                   & $\gg$                   & $\gg$                   & $\gg$                   & $\gg$                       &                        &                        &                          &                          &                        &                        &                        &                          &                          &                          &                        &                          &                          &                          &                          &                          &                          &                          &                        &                          &                          &                          &                        &                          &                          &                         &    \\ 
\cline{3-21}
                                    & \multicolumn{1}{c}{} & \multicolumn{1}{c}{} & \multicolumn{1}{c}{} & \multicolumn{1}{c}{} & \multicolumn{1}{c}{} & \multicolumn{1}{c}{} & \multicolumn{1}{c}{} & \multicolumn{1}{c}{} & \multicolumn{1}{c}{} & \multicolumn{1}{c}{} & \multicolumn{1}{c}{} & \multicolumn{1}{c}{} & \multicolumn{1}{c}{} & \multicolumn{1}{c}{} & \multicolumn{1}{c}{} & \multicolumn{1}{c}{} & \multicolumn{1}{c}{} & \multicolumn{1}{c}{} & \multicolumn{1}{c}{} &                          &                        &                        &                          &                          &                        &                        &                        &                          &                          &                          &                        &                          &                          &                          &                          &                          &                          &                          &                        &                          &                          &                          &                        &                          &                          &                         &    \\ 
\cline{3-48}
\multirow{2}{*}{OCC-W1
  alignment} &                      & A                    & B                    & P                    & $\gg$                   & $\gg$                   & P                    & $\gg$                   & $\gg$                   & D                    & $\gg$                   & E                    & $\gg$                   & P                    & $\gg$                   & E                    & $\gg$                   & L                    & $\gg$                   & \multicolumn{1}{c|}{$\gg$}  & \multicolumn{1}{c|}{K} & \multicolumn{1}{c|}{O} & \multicolumn{1}{c|}{$\gg$}  & \multicolumn{1}{c|}{$\gg$}  & \multicolumn{1}{c|}{G} & \multicolumn{1}{c|}{H} & \multicolumn{1}{c|}{I} & \multicolumn{1}{c|}{$\gg$}  & \multicolumn{1}{c|}{$\gg$}  & \multicolumn{1}{c|}{$\gg$}  & \multicolumn{1}{c|}{E} & \multicolumn{1}{c|}{$\gg$}  & \multicolumn{1}{c|}{$\gg$}  & \multicolumn{1}{c|}{$\gg$}  & \multicolumn{1}{c|}{$\gg$}  & \multicolumn{1}{c|}{$\gg$}  & \multicolumn{1}{c|}{$\gg$}  & \multicolumn{1}{c|}{$\gg$}  & \multicolumn{1}{c|}{I} & \multicolumn{1}{c|}{$\gg$}  & \multicolumn{1}{c|}{$\gg$}  & \multicolumn{1}{c|}{$\gg$}  & \multicolumn{1}{c|}{M} & \multicolumn{1}{c|}{$\gg$}  & \multicolumn{1}{c|}{$\gg$}  & \multicolumn{1}{c|}{F}  & I  \\ 
\cline{3-48}
                                    &                      & A                    & B                    & P                    & $\tau$                  & $\tau$                  & P                    & $\tau$                  & $\tau$                  & D                    & $\tau$                  & E                    & $\tau$                  & P                    & $\tau$                  & E                    & $\tau$                  & L                    & $\tau$                  & \multicolumn{1}{c|}{$\tau$} & \multicolumn{1}{c|}{K} & \multicolumn{1}{c|}{O} & \multicolumn{1}{c|}{$\tau$} & \multicolumn{1}{c|}{$\tau$} & \multicolumn{1}{c|}{G} & \multicolumn{1}{c|}{H} & \multicolumn{1}{c|}{I} & \multicolumn{1}{c|}{$\tau$} & \multicolumn{1}{c|}{$\tau$} & \multicolumn{1}{c|}{$\tau$} & \multicolumn{1}{c|}{E} & \multicolumn{1}{c|}{$\tau$} & \multicolumn{1}{c|}{$\tau$} & \multicolumn{1}{c|}{$\tau$} & \multicolumn{1}{c|}{$\tau$} & \multicolumn{1}{c|}{$\tau$} & \multicolumn{1}{c|}{$\tau$} & \multicolumn{1}{c|}{$\tau$} & \multicolumn{1}{c|}{I} & \multicolumn{1}{c|}{$\tau$} & \multicolumn{1}{c|}{$\tau$} & \multicolumn{1}{c|}{$\tau$} & \multicolumn{1}{c|}{M} & \multicolumn{1}{c|}{$\tau$} & \multicolumn{1}{c|}{$\tau$} & \multicolumn{1}{c|}{$\gg$} & I  \\
\cline{3-48}
\end{tabular}
}
\caption{Example of a trace's prefix-alignment from BPI 2012 log with IM 0.2 threshold.}
\label{table:alignments}
\end{table*}

An example prefix-alignment of the BPI 2012 log is shown in Table~\ref{table:alignments}. The prefix-alignment from the IWS algorithm is much shorter, because the proxy trie does not contain silent transitions. The OCC-W1 prefix-alignment has many model moves on silent transitions, allowing it to find synchronous moves for almost every event in the trace. By convention, the silent transitions are not penalized and have an alignment cost of 0, as they are valid passages through the model. However, it can be argued that while the OCC-W1 alignment has a much lower cost - it has a cost of 1, compared to the IWS cost of 13 - the alignment itself becomes hard to decipher and offers little value to an analyst trying to pinpoint deviations. Thus, such an alignment provides little value in a real-life setting. 
For reference, the single worst cost error in the BPI 2012 log is for a trace which has 170 events. IWS reports a cost of 164, while OCC-W1 reports a cost of 2. However, in the prefix-alignment, OCC-W1 has 537 moves on silent transitions. In total, the output from OCC-W1 across the whole BPI 2012 log for the IM 0.2 model has 76172 silent transitions across the 1000 sample traces. 


Thus to conclude, IWS outperforms OCC-W1 in terms of computation time and is comparable or better in terms of cost error in most cases. In case the model allows for a large variety of behavior, e.g. when dealing with \emph{flower models}, then the cost difference between IWS and OCC-W1 is indeed prominent and for such models OCC-W1 is better suited, especially if calculation time is not a constraint. Importantly, however, it can be argued that the alignments produced by the OCC-W1 for \emph{flower models} are complex to grasp. Furthermore, computing conformance for models which allow any kind of behavior does not seem practical, as it is counterintuitive to the purpose of conformance checking.

\subsection{Stress test}

\begin{figure}
    \centering
    \includegraphics[height=0.3\columnwidth]{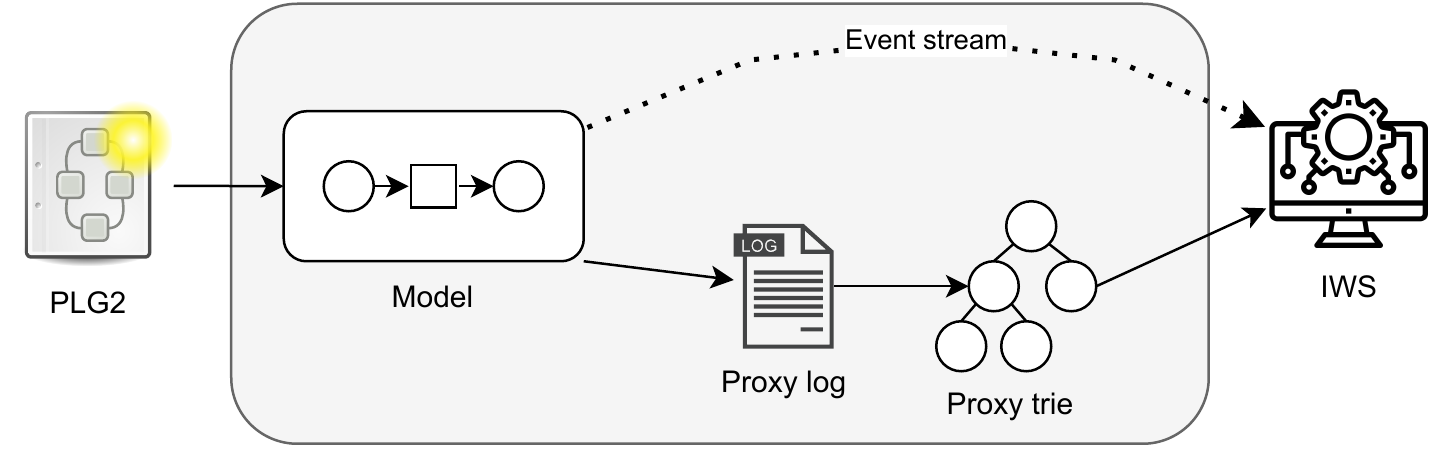}
	
    \caption{Setup for stress testing. Gray rectangle shows the artifacts produced by the PLG2 software.}
    \label{fig:approach_PLG}
\end{figure}

In order to test the algorithm for speed and memory consumption, the PLG2 software \cite{burattin2016plg2} was used to simulate three process models of various sizes. PLG2 was further used to simulate a proxy log from these models, and to stream events to the socket while adding different degrees of noise to the event stream. An overview of the approach can be seen in Figure~\ref{fig:approach_PLG}. The memory consumption was measured using the tool VisualVM \cite{sedlacek2017visualvm}.

\begin{table}[]
\centering
\scalebox{0.8}{
\begin{tabular}{ccccc}
\hline
\textbf{\begin{tabular}[c]{@{}c@{}}Model \\ type\end{tabular}} & \textbf{Activities} & \textbf{\begin{tabular}[c]{@{}c@{}}Traces in \\ proxy log\end{tabular}} & \textbf{Trie nodes} & \textbf{\begin{tabular}[c]{@{}c@{}}Trie construction \\ time (ms)\end{tabular}} \\ \hline
Small                                                          & 16                  & 256                                                                     & 841                 & 1267                                                                            \\
Medium                                                         & 153                 & 23409                                                                   & 713640              & 19934                                                                           \\
Large                                                          & 471                 & 25000                                                                   & 760343              & 101960                                                                          \\ \hline
\end{tabular}
}
\caption{Simulated models description}
\label{table:simdescription}
\end{table}

A description of the process models generated by PLG2 and the resulting tries is in Table \ref{table:simdescription}. The number of unique activities illustrate the complexity level of each of the models. The proxy logs were simulated based on a multiple of the activities in the model, except for the \emph{Large} model, where the simulation was stopped at 25000 traces due to long execution time of the simulation process. 

The event streams were created with three alternating default settings in PLG2: no noise, low noise (5\%), and high noise (10\%). The noise setting indicates whether the event stream can be expected to be conforming to the process model or not. With no noise, all of the activities in the stream are conforming exactly to the process model. Simulating noise generates traces which are non-conforming to the model, e.g. by introducing behavior not explained by the model, or skipping behavior that is required by the model. Each of the event streams was kept running for 5 minutes. Table~\ref{table:stresstest} shows the results from the executions per model and noise type.  

\begin{table}[]
\centering
\scalebox{0.6}{
\begin{tabular}{@{}cccccccc@{}}
\toprule
\begin{tabular}[c]{@{}c@{}}Model \\ type\end{tabular} & \begin{tabular}[c]{@{}c@{}}Noise \\ level\end{tabular} & Events & \begin{tabular}[c]{@{}c@{}}Computation \\ time (ms)\end{tabular} & \begin{tabular}[c]{@{}c@{}}Idle \\ time (ms)\end{tabular} & \begin{tabular}[c]{@{}c@{}}Computation \\ ms/event\end{tabular} & \begin{tabular}[c]{@{}c@{}}CPU \\ usage\end{tabular} & \begin{tabular}[c]{@{}c@{}}Memory \\ usage\end{tabular} \\ \midrule
                                                      & None                                                   & 50184  & 49954                                                            & 250045                                                    & 1.0                                                             & \textless{}5\%                                       & 10-80MB                                                 \\
Small                                                 & Low                                                    & 51088  & 48813                                                            & 251144                                                    & 1.0                                                             & \textless{}5\%                                       & 20-80MB                                                 \\
                                                      & High                                                   & 47666  & 43553                                                            & 256408                                                    & 0.9                                                             & \textless{}5\%                                       & 10-80MB                                                 \\ \cmidrule(l){2-8} 
                                                      & None                                                   & 72481  & 152631                                                           & 147344                                                    & 2.1                                                             & \textless{}10\%                                      & 1000-1600MB                                             \\
Medium                                                & Low                                                    & 70692  & 145348                                                           & 154596                                                    & 2.1                                                             & \textless{}10\%                                      & 1000-1600MB                                             \\
                                                      & High                                                   & 69590  & 160139                                                           & 139782                                                    & 2.3                                                             & \textless{}10\%                                      & 600-1200MB                                              \\ \cmidrule(l){2-8} 
                                                      & None                                                   & 82025  & 183702                                                           & 115658                                                    & 2.2                                                             & \textless{}15\%                                      & 1500-2000MB                                             \\
Large                                                 & Low                                                    & 81493  & 182374                                                           & 117623                                                    & 2.2                                                             & \textless{}15\%                                      & 1500-2000MB                                             \\
                                                      & High                                                   & 76511  & 177086                                                           & 122283                                                    & 2.3                                                             & \textless{}15\%                                      & 1500-2000MB                                             \\ \bottomrule
\end{tabular}
}
\caption{Stress test results}
\label{table:stresstest}
\end{table}

The algorithm was successfully able to keep up with the stream. For the small model, the algorithm was idle for $\frac{5}{6}$ of the time, signaling that higher throughput could have been achieved. For medium and large models, the algorithm was idle for about half the time or less.

As expected, the experiments using the small model had the least strain on the CPU and memory, with less than 5\% of CPU utilized by the Java application running the algorithm, and memory usage varying between 10-80 MB during the execution. For medium and large models, the CPU usage was still relatively low, being less than 10\% and less than 15\%, respectively. Memory usage was much higher - likely due to a higher amount of nodes in the medium and large tries and longer traces, which lead to more states being kept in memory. For the medium model, the memory usage was between 600-1600 MB, while for the large model, the memory usage was between 1500-2000 MB for each of the event streams.

It was expected that higher noise levels would increase the computation time, as the algorithm needs to make more computationally costly non-synchronous moves and state space exploration. The effect is visible for medium and large models with a high noise level setting, as the computation time per event slightly increases, but the increase is not drastic. This indicates that the IWS algorithm is able to handle event streams in a relatively equal manner despite the level of deviations occuring in the stream. 

The CPU and memory usage results show that the algorithm is usable also on machines with low-end hardware. Importantly, the current experiments did not incorporate mechanisms to remove the processed traces from memory. While this would prove unfeasible in real life, options for handling the memory and limiting the cases kept in memory have been researched separately in the literature  \cite{zaman2021efficient} and it is out of scope for this paper. Suffice to say, the algorithm can be extended to support such functionality that would allow setting a decision-point when to remove a case from memory, or defining a logic when memory would be flushed to disk. 

Ultimately, the algorithm displays very fast processing of event streams with a low strain on memory. Thus, the algorithm would be applicable for real-life, streaming online conformance checking.

\section{Conclusion} \label{sec:conclusion}
In this paper, a new approximate algorithm (IWS) for calculating prefix-alignments in online conformance checking was introduced. The algorithm uses a trie as the underlying structure for holding the model behaviour. \emph{State buffer} is used as a way to keep track of seen traces, \emph{decay time} is used for releasing states from the buffer, and \emph{look-ahead limit} is used for optimizing possible model moves. 

The IWS algorithm was compared against the current state-of-the-art (OCC-W1) using a variety of synthetic and real-life datasets. IWS outperformed OCC-W1 in all instances in terms of computation time. In some cases, IWS produced an output in 8-13 seconds, while OCC-W1 failed to finish within an hour, indicating roughly 3 orders of magnitude computation time difference. At the same time, the IWS achieved comparable cost error and for over a quarter of the datasets even achieved a lower error cost than OCC-W1. 

The IWS algorithm was shown to have poor error cost for process models with low precision - \emph{flower models}. However, the alignments produced by OCC-W1 for such models are hard to decipher. Furthermore, conformance checking on models allowing any kind of behavior is arguably not sensible.

The algorithm was also stress tested using a fast-paced event stream. It was shown to have fast processing time while remaining memory efficient. 

Future work directions are manifold. 

The algorithm expects as input a trie, which is constructed from a proxy log derived from the model. This implies several dependencies on external factors, such as the quality of the method simulating the proxy log from the model. Thus, one research direction would be to generate a trie directly from an existing process model. 

Another direction is to incorporate behavioral metrics into the prefix-alignment calculation. This would also lead to potential solutions for handling warm-starting scenarios - i.e., cases where the stream is ongoing before an online conformance checking method is applied.

Furthermore, current research in online conformance checking has not touched upon the possibility of stream imperfections. In real-life settings, imperfections such as out-of-order events are likely to occur. IWS, utilizing a discounted Decay Time, could be a solution to handle stream imperfections. 

Finally, from a practical point of view, the IWS algorithm could be extended to existing frameworks dealing with event streaming, such as the Beamline Framework\footnote{\url{https://www.beamline.cloud}} which utilizes Apache Flink. This would allow for truly scalable and performant online conformance checking using prefix-alignments to be practiced.

\section*{Acknowledgement}
This work was supported by the European Social Fund via "ICT programme" measure, the European Regional Development Fund, and the programme Mobilitas Pluss (2014-2020.4.01.16-0024).

\bibliographystyle{splncs04}
\bibliography{bibliography}

\end{document}